\documentclass[10pt]{iopart}

\usepackage{iopams}  

\usepackage[superscript,biblabel]{cite}
\usepackage[colorlinks]{hyperref}\hypersetup{
     colorlinks   = true,
     citecolor    = blue
}
\usepackage{hyperref}
\hypersetup{
    colorlinks=true,
    linkcolor=blue,
    filecolor=magenta,      
    urlcolor=cyan,
}
\usepackage[normalem]{ulem}
\usepackage[toc,page]{appendix}
\usepackage[utf8]{inputenc}
\usepackage[T1]{fontenc}
\usepackage{lmodern}
\usepackage[english]{babel}
\usepackage{color}
\usepackage{graphics}
\usepackage{epsfig}
\usepackage{float}
\usepackage{array}
\usepackage{bm}
\usepackage{footnote}
\usepackage{gensymb}
\usepackage[symbol]{footmisc}
\usepackage{array}
\usepackage{makecell}
\usepackage{tabularx}
\usepackage{pdfpages}

\newcolumntype{C}[1]{>{\centering\arraybackslash$}p{#1}<{$}}
\usepackage{fancyhdr}
\usepackage{lastpage}

\usepackage[heightrounded]{geometry}
\usepackage{iopams} 

\begin{document}

\setlength{\headwidth}{\textwidth}
\pagestyle{fancyplain}
\lhead{ \fancyplain{}{Nucl. Fusion XX (2024) XXXXXX} }
\rhead{ \fancyplain{}{A. Kumar \it et  al} }
\setlength{\headwidth}{\textwidth}
\addtolength{\headwidth}{\marginparsep}
\addtolength{\headwidth}{\marginparwidth}

\title[]{Integrated modeling of RF-Induced Tungsten Erosion at ICRH Antenna Structures in the WEST Tokamak\mbox{*}}

\author{A. Kumar$^{1}$$^\dagger$, W. Tierens$^1$, T. Younkin$^1$, C. Johnson$^1$, C. Klepper$^1$, A. Diaw$^1$, J. Lore$^1$,  A. Grosjean$^2$, G. Urbanczyk$^3$, J. Hillairet$^3$,  P. Tamain$^3$, L. Colas$^3$, C. Guillemaut$^3$   D. Curreli$^4$, S. Shiraiwa$^5$, N. Bertelli$^5$, and the WEST Team$^a$  }

\address{$^1$Oak Ridge National Laboratory, 1 Bethel Valley Road, Oak Ridge, TN-37831, USA}
\address{$^2$University of Tennessee, Knoxville, TN 37996, USA }
\address{$^3$CEA, IRFM, F-13108 Saint-Paul-lez-Durance, France }
\address{$^4$University of Illinois at Urbana Champaign, Urbana, IL 61801, USA }
\address{$^5$Princeton Plasma Physics Laboratory, Princeton, NJ 08536, USA }
\address{$^a$see http://west.cea.fr/WESTteam }

\ead{$^\dagger$kumara@ornl.gov}


\begin{abstract}

This paper introduces STRIPE (Simulated Transport of RF Impurity Production and Emission), an advanced modeling framework designed to analyze material erosion and the global transport of eroded impurities originating from radio-frequency (RF) antenna structures in magnetic confinement fusion devices. STRIPE integrates multiple computational tools, each addressing different levels of physics fidelity: SolEdge3x for scrape-off-layer plasma profiles, COMSOL for 3D RF rectified voltage fields, RustBCA code for erosion yields and surface interactions, and GITR for 3D ion energy-angle distributions and global impurity transport.  The framework is applied to an ion cyclotron RF heated, L-mode discharge  \#57877 in the WEST Tokamak, where it predicts a tenfold increase in tungsten erosion at RF antenna limiters under RF-sheath rectification conditions, compared to cases with only a thermal sheath. Highly charged oxygen ions ($\rm O^{6^+}$ and higher) emerge as dominant contributors to tungsten sputtering at the antenna limiters. To verify model accuracy, a synthetic diagnostic tool based on inverse photon efficiency or \rm S/XB coefficients from the ColRadPy- collisional radiative model enables direct comparisons between simulation results and experimental spectroscopic data. Model  predictions, assuming plasma composition of 1\% oxygen and 99\% deuterium,  align closely with measured neutral tungsten (W-I) spectroscopic data for the discharge  \#57877, validating the framework's accuracy. Currently, the STRIPE framework is being extended to investigate plasma-material interactions in other RF-heated linear and toroidal devices, offering valuable insights for RF antenna design, impurity control, and performance optimization in future fusion reactors.

\footnotetext{This manuscript has been authored by UT-Battelle, LLC, under contract DE-AC05-00OR22725 with the US Department of Energy (DOE). The US government retains and the publisher, by accepting the article for publication, acknowledges that the US government retains a nonexclusive, paid-up, irrevocable, worldwide license to publish or reproduce the published form of this manuscript, or allow others to do so, for US government purposes. DOE will provide public access to these results of federally sponsored research in accordance with the DOE Public Access Plan (http://energy.gov/downloads/doe-public-access-plan).}
\end{abstract}
%
%
%
%
\ioptwocol
\section{Introduction: Context and objectives}
\label{sec:1}

High-power radio-frequency (RF) waves are a crucial technique for plasma heating in magnetic confinement fusion (MCF) devices, enabling efficient energy transfer to achieve fusion conditions. RF heating methods have proven highly effective, as demonstrated by the 2022 deuterium-tritium (DT) campaign at the Joint European Torus (JET), where ion-cyclotron RF heating (ICRH) contributed ~20\% of the record-setting 59 MJ fusion energy output\cite{Maslov_2023}.  As fusion research progresses toward practical energy generation, upcoming MCF-based experimental devices like SPARC \cite{Creely_2020, Rodriguez-Fernandez_2022} and  Material Plasma Exposure eXperiment (MPEX) \cite{Rapp_2020, Kumar_2023} plan to rely heavily, if not exclusively, on RF-based plasma heating. This trend underscores the importance of understanding and addressing the unique challenges posed by RF heating and current drive in high-performance fusion environments.

One significant challenge to achieving reliable RF heating and current drive in next-generation fusion devices, including ITER, is the interaction of RF waves with plasma sheaths near material surfaces such as RF antennas structures. RF rectified potentials, or RF sheaths \cite{myra2021tutorial}, form on these surfaces with amplitudes that can reach several hundred volts in current MCF devices \cite{Tierens_2024} and are projected to reach up to kilovolts levels in future devices like ITER, where slow-wave resonances are anticipated to play a critical role \cite{tierens2023resonance}. These high RF potentials accelerate plasma fuel ions, such as hydrogen ($\rm H^+$), deuterium ($\rm D^+$), and light impurity ions (e.g., oxygen, boron, neon etc.), towards the antenna surfaces, potentially causing substantial material erosion. An additional complication is self-sputtering, in which sputtered materials, primarily from the antenna surface, return to the surface, exacerbating ongoing erosion and altering the surface morphology.

This cycle of erosion and re-deposition introduces impurities into the scrape-off-layer (SOL) neutral gas and plasma environment, posing a risk of impurities migrating into the plasma core. This issue is particularly problematic during high-confinement mode (H-mode) operations, as longer plasma confinement times increase impurity retention in the core, potentially degrading plasma performance \cite{Angioni:2021, Angioni:2014 }. Recent studies have observed substantial material erosion at RF antenna structures and subsequent transport of these impurities towards the bulk plasma during helicon and ICRH in both linear \cite{Beers1, Beers2} and toroidal MCF devices \cite{Klepper2022}. In past, several efforts to mitigate impurity generation have been explored, including (1) wall boronization at TEXTOR \cite{waelbroeck1989influence}, (2) toroidal phasing of antenna straps at JET \cite{bures1988modification} and ASDEX Upgrade (AUG) \cite{bobkov2016making, bobkov2019impact}, and (3) the use of field-aligned ICRH antenna arrays at Alcator C-Mod \cite{wukitch2013characterization}. However, the fusion community recognizes the need for more advanced strategies and a deeper understanding of RF-enhanced impurity production to minimize the impact on plasma performance in future reactor-grade fusion environments.

To address these challenges, this study introduces STRIPE (Simulated Transport of RF Impurity Production and Emission), a comprehensive modeling framework for studying RF-induced PMI at ICRH antenna structures in fusion devices. STRIPE integrates advanced computational tools across various physics fidelity levels (see Section \ref{sec:2}), enabling detailed simulations of impurity sources originating from ICRH antenna structures in MCF experiments. The results of STRIPE simulations are directly compared to observed tungsten (W) erosion at ICRH antenna limiters in the WEST Tokamak during a targeted ICRH L-mode discharge \#57877, with particular focus on quantifying erosion and impurity transport processes.

WEST, with its all-W plasma-facing components (PFCs), provides an optimal environment for systematically studying impurity sources from both the divertor and main-chamber components, including ICRH antenna limiters. The device’s long-pulse operational capability (up to 1000s) also allows for extended investigations of PMI under long-duration tokamak conditions. Recent research in WEST \cite{Klepper2022} has highlighted significant W impurity generation from main-chamber ICRH antenna structures, although limitations in computational tools at the time prevented consideration of self-sputtering effects and net W erosion at the antenna limiters. The STRIPE framework addresses these limitations by providing a high-fidelity simulation platform capable of modeling these complex interactions on RF environment. This study on RF induced PMI at the WEST ICRH antenna limiters is especially relevant to ITER, which has recently adopted an all-W PFC design, underscoring the need for accurate models to predict and manage impurity production in high-power, reactor-grade fusion devices.

This paper is organized as follows: Section \ref{sec:2} introduces the STRIPE modeling framework, detailing the computational tools and methodologies used to simulate W erosion and impurity transport in RF antenna structures. Section \ref{sec:3} describes the modeling of the WEST background plasma, providing the plasma conditions and profiles necessary for STRIPE inputs. Section \ref{sec:4} discusses additional inputs required for STRIPE, including RF sheath boundary conditions and material erosion yield calculations. Section \ref{sec:5} presents the synthetic diagnostic method developed to convert simulated W erosion flux into photon emission, allowing comparison with experimental data. Section \ref{sec:6} presents the main results on W erosion at the WEST ICRH antenna limiters, comparing thermal and RF sheath cases and validating model predictions with experimental observations. The discussion in Section \ref{sec:7} explores the implications of these findings for PMI in fusion devices, potential improvements in the STRIPE framework, and future directions. Finally, summary of this work is presented in Section \ref{sec:8}.

\section{Simulated Transport of RF Impurity Production and Emission—STRIPE Framework} 
\label{sec:2}

To accurately estimate material erosion and the transport of sputtered impurities at ICRH antenna structures in fusion devices, it is essential to model physical processes across multiple temporal and spatial scales. This complexity necessitates an integrated framework, which the STRIPE model addresses by combining several specialized computational tools, each capturing different aspects of the physics involved. The key components of this modeling framework are outlined below:

\begin{itemize}
    \item \textbf{Background Plasma Profiles:} A primary input to the STRIPE framework is the plasma background, including electron density,  temperature, and ion fluxes. These parameters are obtained using fluid-based SOL plasma transport codes, such as SOLPS \cite{Schneider:2006, Islam:2023} and SolEdge3x \cite{Bufferand_2015, Ciraolo_2019}. For this study, the WEST SOL plasma background is modeled with SolEdge3x, which provides profiles of both plasma and light impurities (e.g., oxygen) across the SOL and far-SOL regions on an extended computational grid which extends radially to the main-chamber and antenna PFCs. Using extended-grid SOL transport codes like SolEdge3x is crucial, particularly when experimental diagnostics are unavailable to characterize plasma and impurity profiles near the antenna structures, where PMI are most intense. Details of these simulations are provided in Section \ref{sec:3}.
    
    \item \textbf{RF Sheath Boundary Conditions and RF Wave Propagation:} Following established methods presented in references \citen{Beers1} and \citen{Beers2}, RF sheaths are simulated as thin dielectric layers mimicking sheath behavior using COMSOL multi-physics package, as described also in a recent study \cite{Tierens_2024}. This approach allows for the modeling of RF rectified sheath potentials and the propagation of RF waves in a cold plasma environment, establishing essential boundary conditions for the STRIPE framework.
    
    \item \textbf{CAD Defeaturing of the ICRH Antenna Structures:} Realistic geometry representation is crucial for STRIPE modeling, particularly in finite-element simulations requiring precise computational geometry of RF antenna structures. However, engineering CAD models must undergo "defeaturing"—removing non-essential details and refining the mesh in critical areas—to balance computational efficiency and accuracy.
    
    \item \textbf{Material Erosion Yield and Surface Interaction Physics:} The RustBCA code \cite{RustBCA, Drobny:2023} is utilized to simulate ion-surface interactions at the atomic scale, delivering data on sputtering and reflection yields over a wide spectrum of incident ion energies (ranging from a few eV to 10,000 eV) and angles (0–90°). This tool focuses on ion-surface interactions, particularly sputtering, reflection, and implantation, which are difficult to model analytically due to the limitations of empirical or semi-empirical formulas in addressing complex geometries, multi-component surfaces, or dynamic plasma-material coupling. Using the Binary Collision Approximation (BCA) as a computationally efficient alternative to n-body molecular dynamics methods, RustBCA enables accurate and scalable simulations. While molecular dynamics provides more detailed insights, it is computationally infeasible for intricate RF antenna geometries, making RustBCA a practical and efficient solution for high-fidelity surface interaction modeling within STRIPE.
    
    \item \textbf{Calculation of Ion Energy-Angle Distributions (IEADs) at Material Surfaces:} A critical component of this modeling framework is resolving the complex physics within the plasma sheath adjacent to material surfaces, where RF wave rectification significantly affects plasma behavior. To capture both thermal and non-thermal plasma sheath effects, a fully kinetic treatment is desirable, which could be achieved using Particle-In-Cell (PIC) codes like HPIC \cite{KHAZIEV201887}. However, to manage computational costs in the vicinity of complex RF antenna geometries, the 3D tracer-particle code - GITR \cite{GITR} is used to calculate 3D energy-angle distributions  of ions impacting material surfaces. This approach incorporates interactions within the plasma sheath, considering the full 3D geometry of the antenna structures, as described in Section \ref{sec:4}.
\end{itemize}

\begin{figure}
    \includegraphics[width=\linewidth]{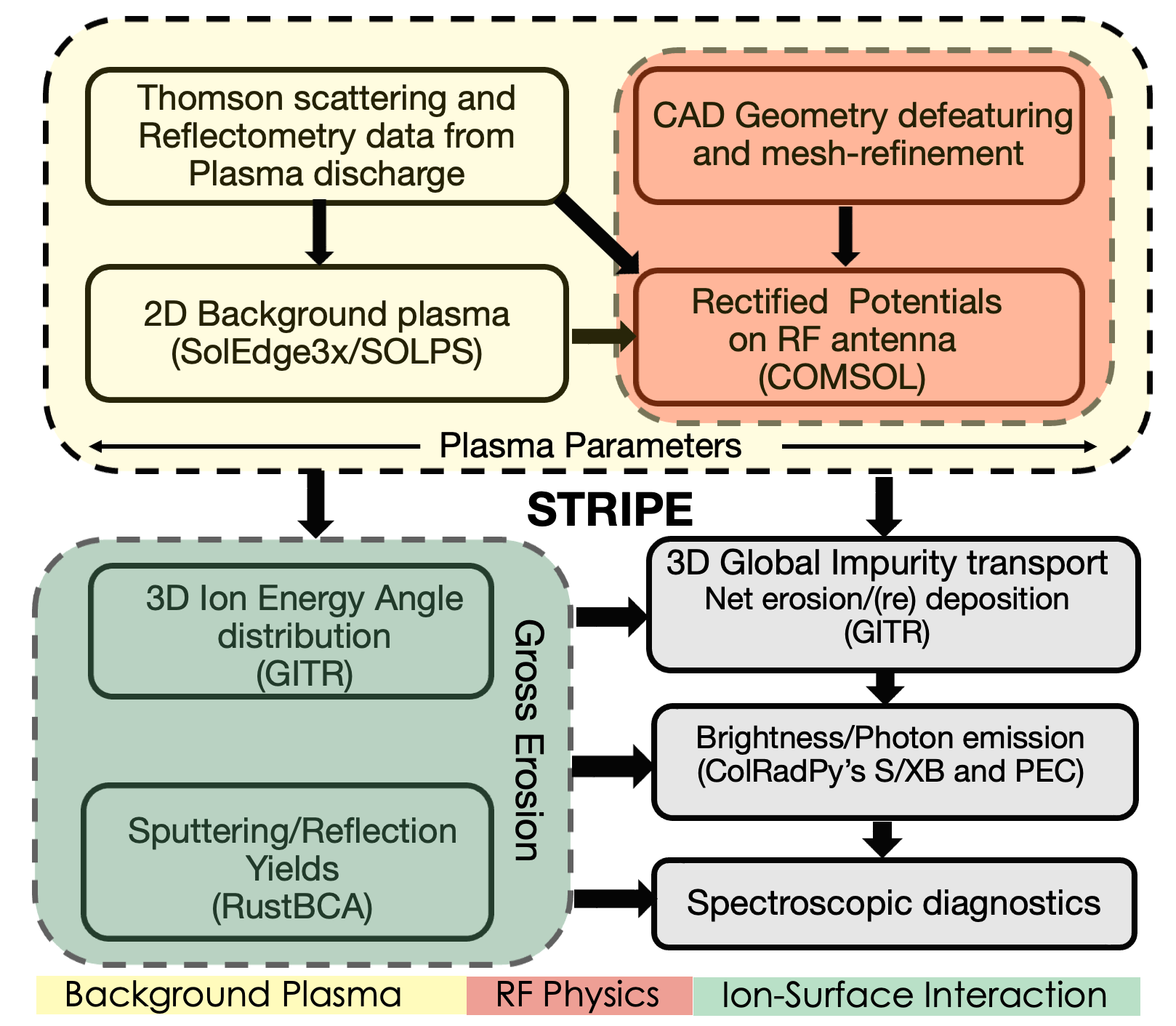}
    \caption{STRIPE framework workflow for modeling W erosion and impurity transport at RF antenna structures. Components include SolEdge3x for SOL plasma profiles, COMSOL for RF sheath potentials, RustBCA for sputtering yield calculations, GITR for particle tracking, and ColRadPy for synthetic diagnostic conversion to brightness, $\rm I_\phi$ ({\emph{i.e.}} photon flux) for experimental comparison.}
    \label{fig:1}
\end{figure}

Using these inputs, the Global Impurity TRansport (GITR)- a 3D Monte Carlo particle tracker code \cite{GITR} predicts material erosion on the RF antenna surfaces. To facilitate direct comparison with experimental data, synthetic diagnostics are employed, incorporating inverse photon efficiency (\rm S/XB) and photon emissivity coefficients (PEC) from the ColRadPy collisional radiative model \cite{curt2019}. This diagnostic process translates impurity flux data into brightness or photon flux $\rm I_\phi$, producing line emission results that can be directly compared with experimental spectroscopic measurements for validation. The complete STRIPE framework workflow is illustrated in Fig. \ref{fig:1}.

The predictive accuracy of material erosion and impurity transport within STRIPE as also discussed later in Section \ref{subssec:74}, heavily relies on the precision of the input parameters and models outlined here. In this study, we apply STRIPE to simulate RF-induced W erosion at the ICRH antenna structures within the toroidal geometry of the WEST tokamak, as outlined in Section \ref{sec:3}.

\section{Modeling of the WEST Background Plasma}
\label{sec:3}
The WEST tokamak features a fully metallic environment designed for steady-state operation, with key components optimized for plasma-material interaction studies relevant to ITER. A critical aspect of these studies is the configuration of the radio frequency (RF) heating systems and diagnostics deployed within the vessel.

\begin{figure}
    \includegraphics[width=\linewidth]{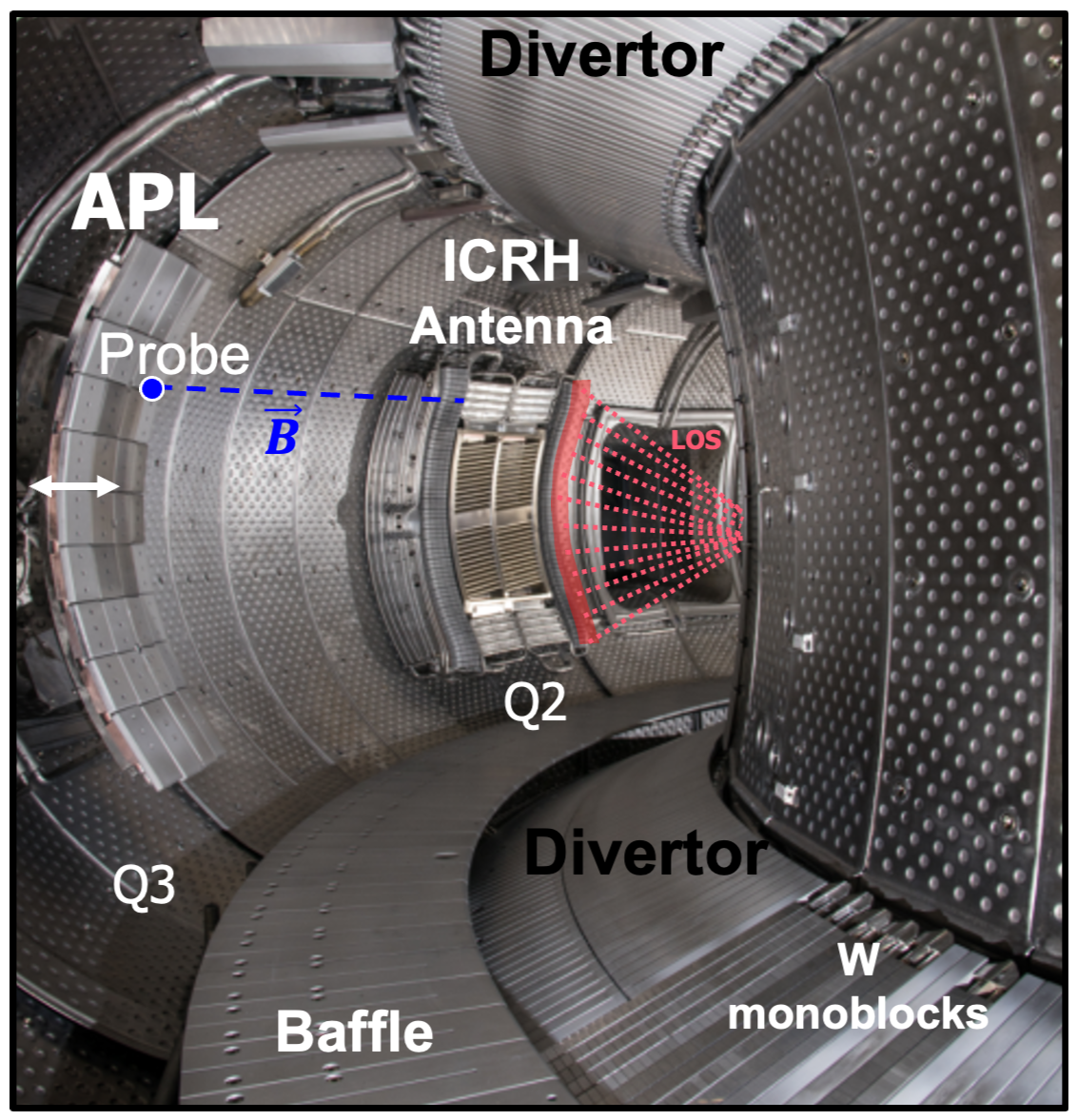}
    \caption{View of the WEST inner vessel showing a magnetic field line (blue dashed) connecting the upper Pecker probe on the Antenna Protection Limiter (APL) to the upper left corner of the Q2 ICRF antenna. The image highlights the visible spectrometer’s lines of sight (red dashed) embedded in the high-field side wall, focusing on the Q2 ICRF antenna limiters. Also visible are the baffle and the divertor, which consists primarily of W-coated graphite tiles, with 12 tiles featuring ITER-like W monoblocks.}
    \label{fig:2a}
\end{figure}

Figure \ref{fig:2a} provides a detailed view of the WEST inner vessel, showcasing a magnetic field line (blue dashed line) that connects the upper Pecker probe on the Antenna Protection Limiter (APL) to the upper left corner of the Q2 ICRF antenna. The visible spectroscopy diagnostics (red dashed lines) are embedded in the high-field side wall, specifically targeting the Q2 ICRF antenna limiters. The image also highlights the baffle structure and the divertor region. The divertor primarily consists of W-coated graphite tiles, with 12 tiles equipped with ITER-like W monoblocks to simulate realistic plasma-facing conditions. This configuration supports detailed diagnostics of plasma-material interactions and impurity generation under high-power RF heating scenarios relevant to ITER.

\begin{figure}
    \includegraphics[width=\linewidth]{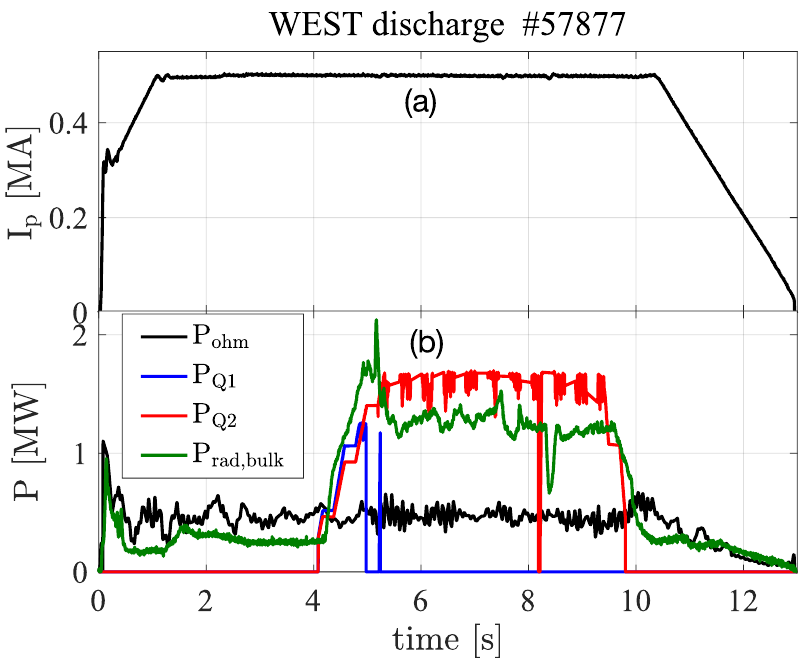}
    \caption{(a) Plasma current profile (\(I_p\)) for WEST L-mode discharge \#57877, showing the current ramp-up phase until \(t=1\)~s, followed by a flattop at \(I_p=0.5\)~MA maintained until \(t=10\)~s. (b) Power input profiles, including \(\rm P_{ohm}\) (black curve) for ohmic heating, \(\rm P_{Q1}\) (red curve) and \(\rm P_{Q2}\) (blue curve) for ICRH power applied through Q1 and Q2 RF antennas, respectively, and the total radiated power (\(\rm P_{rad, bulk}\)). The transition from the ohmic phase (\(t=0{-}5\)~s) to the ICRH phase (\(\rm t=5-10\)~s) is marked by a power increase from \(\rm P_{ohm}=0.6\)~MW to a total input power of approximately \(2.3\)~MW, with \(\rm P_{Q2}=1.7\)~MW during the ICRH phase. }
 
    \label{fig:2}
\end{figure}

To simulate W erosion at the ICRH antenna structures, this study utilizes experimental conditions from WEST L-mode discharge \#57877 (Fig.~\ref{fig:2}). This discharge serves as a representative WEST plasma scenario, beginning with an ohmic heating phase from \(\rm t=0\)~s to \(\rm t=5\)~s, followed by ion cyclotron resonance heating (ICRH) from \(\rm t=5\)~s to \(\rm t=10\)~s.

Figure~\ref{fig:2} presents (a) the plasma current profile, \(\rm I_p\), and (b) the power input profiles: \(\rm P_{ohm}\) (black curve), representing the power applied for ohmic heating; \(\rm P_{Q1}\) (red curve) and \(\rm P_{Q2}\) (blue curve), showing the externally applied ICRH power from the Q1 and Q2 RF antennas, respectively; and the total radiated power (\(\rm P_{rad, bulk}\)). These profiles provide an overview of the heating stages and energy distribution. In this discharge, ICRH power is applied exclusively through the Q2 antenna.

As depicted in Fig.~\ref{fig:2}, the plasma current reaches \(\rm I_p=0.5\)~MA by \(\rm t=1\)~s and remains at a flattop until \(\rm t=10\)~s. During the ohmic phase, the applied power is approximately \(\rm P_{ohm}=0.6\)~MW, which increases to a total input power, $\rm P_{in}=P_{Q1}+P_{Q2}+P_{ohm}=2.3$~MW  with the activation of ICRH  through the Q2 antenna (\(\rm P_{Q2}=1.7\)~MW). For the simulations, two specific time points are analyzed: \(\rm t=3\)~s for the ohmic phase and \(\rm t=8\)~s for the ICRH phase. These will henceforth be referred to as the ohmic phase and the ICRH phase, respectively, in this paper.

The plasma simulations for both phases are conducted using the SolEdge-EIRENE package, which combines the SolEdge multi-fluid plasma code \cite{Bufferand_2015} with the EIRENE Monte Carlo neutral particle solver \cite{Reiter_2005}. This study uses the latest version, SolEdge3x \cite{Bufferand_2019, Bufferand_2021}, configured for 2D simulations with an axisymmetric wall assumption.

In present, the plasma composition is set to 99\% $\rm D^+$ and 1\% oxygen, where oxygen serves as a proxy for all light impurities present during the WEST discharge. The oxygen density in the SolEdge3x simulations is fixed at 1\% and initialized from the core boundary. This concentration effectively models impurity behavior and radiative losses, though in WEST experiments it typically ranges between 0.1\% and 10\% \cite{Ciraolo_2019, Klepper2022}. Subsequent sections will demonstrate that W erosion estimates based on this 1\% oxygen concentration agree well with experimental data.

To account for radiative losses in the SOL, the net power crossing the separatrix ($\rm P_{sep}$) was adjusted from the experimentally measured input power, with values of $\rm P_{sep} = 0.4$~MW for the ohmic phase and $\rm P_{sep} = 1.0$~MW for the ICRH phase.

The simulation setup includes a gas puff at the outer midplane (OMP)  at $z=0$~m, calibrated to match $\rm n_{e^-}$ measurements from  the OMP reflectometer. A “transport mode” was applied to approximate turbulence, using fixed diffusive transport coefficients to represent perpendicular  transport. The particle and momentum transport coefficients were set to $\rm D = \nu = 0.4 \, \rm{m}^2/\rm{s}$, typical for WEST L-mode discharges, with ion and electron energy transport coefficients of $\rm \chi_i = \chi_e = 1.0 \, \rm{m}^2/\rm{s}$ \cite{Ciraolo_2019}. A recycling coefficient of $\rm R = 0.99$ was applied at the wall, and a pump albedo coefficient of $\rm R_{\rm pump} = 0.95$ was used, consistent with the WEST pumping system. Drifts are deactivated in these simulations.

\begin{figure*}
\includegraphics[width=\linewidth]{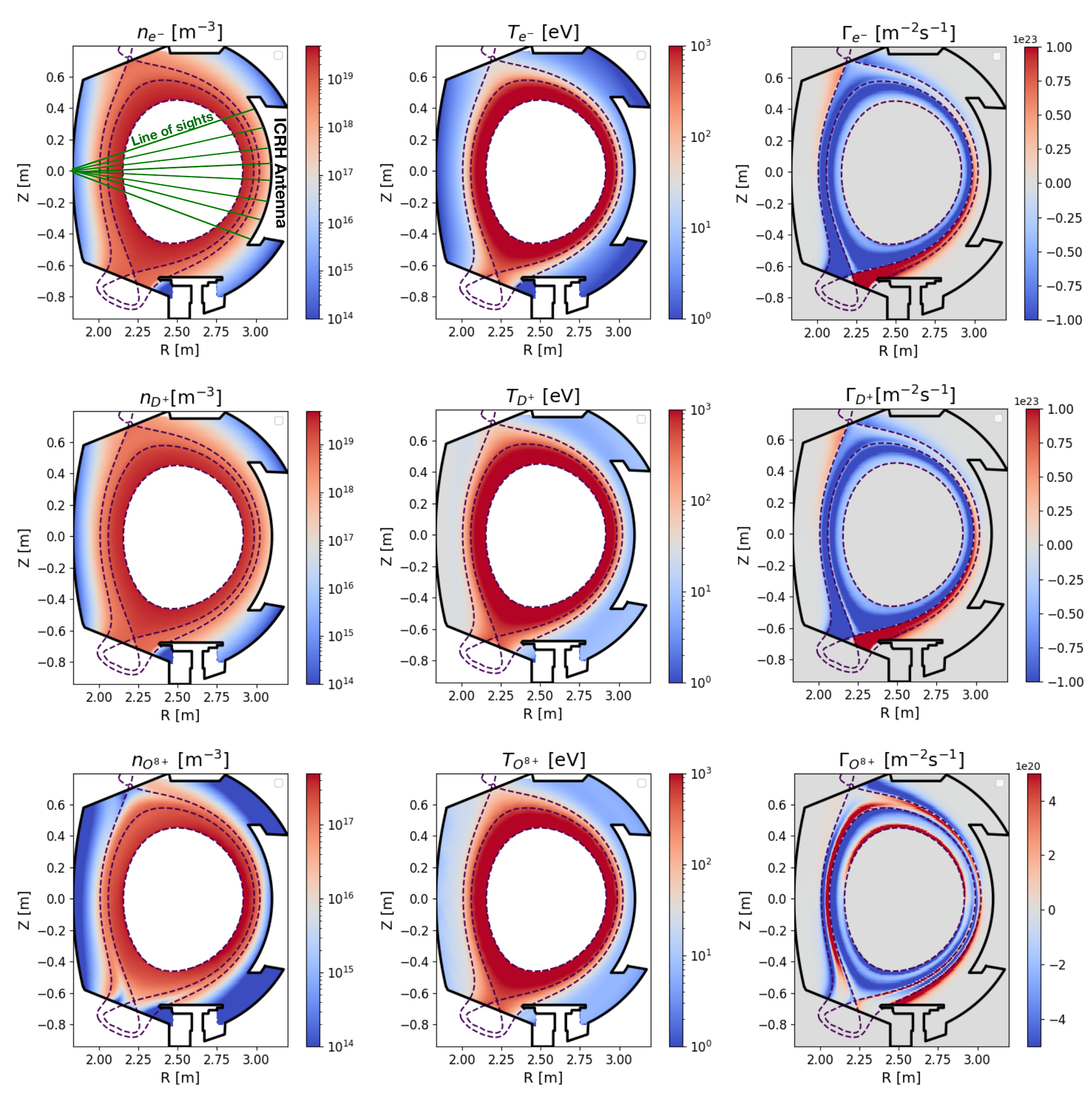}
\caption{Plasma profiles obtained from SolEdge3x simulation with $\rm P_{\rm sep}=1.0$~MW are presented for various species: (a) electron density ($\rm n_{e^-}$), (b) electron temperature ($\rm T_{e^-}$), (c) electron particle flux ($\rm \Gamma_{e^-}$), (d) $\rm D^+$ ion density ($\rm n_{D^+}$), (e) $\rm D^+$ ion temperature ($\rm T_{D^+}$), (f) $\rm D^+$ ion flux ($\rm \Gamma_{D^+}$), (g) oxygen-$\rm 8+$ ion density ($\rm n_{O^{8^+}}$), (h) oxygen-$\rm 8^+$ ion temperature ($\rm T_{O^{8^+}}$), and (i) oxygen-$\rm 8^+$ ion flux ($\rm \Gamma_{O^{8^+}}$). The purple dashed traces represent the inner core boundary, separatrix, and LCFS, while the thick black line indicates the outer wall of the WEST Tokamak. $\rm n_{O^{8^+}}$ is approximately 1\% of $\rm n_{e^-}$. The green lines in the top-left subplot illustrate the line of sight used for measuring line emissions at the IRCH antenna limiters in WEST experiments.} 
\label{fig:4}
\end{figure*}

Figure~\ref{fig:4} displays the simulated background plasma profiles for multiple species, offering insight into the plasma conditions surrounding the ICRH antenna limiters. These profiles include key parameters such as electron density ($\rm n_{e^-}$), temperature ($\rm T_{e^-}$), and flux ($\rm \Gamma_{e^-}$); $\rm D^+$ ion density ($\rm n_{D^+}$), temperature ($\rm T_{D^+}$), and flux ($\rm \Gamma_{D^+}$); and high-charge-state oxygen $\rm O^{8^+}$) ion density ($\rm n_{O^{8^+}}$), temperature  ($\rm T_{O^{8^+}}$), and flux ($\rm \Gamma_{O^{8^+}}$). The green lines in the top-left subplot illustrate the line of sight used for measuring line emissions at the IRCH antenna limiters in WEST experiments. 

Each subplot provides 2D-spatial distributions of these quantities, essential for understanding how plasma conditions vary near the antenna. The purple dashed lines on all the subplots in Fig. \ref{fig:4} show critical magnetic boundaries including inner core boundary, separatrix and LCFS. The outer wall of WEST is indicated by the thick black line. These profiles serve as the background plasma conditions for STRIPE simulations, particularly informing the modeling of PMI at the antenna limiters. 

Fractional abundance analysis for oxygen ions under similar WEST plasma conditions, as presented in reference \citen{Klepper2022}, indicate that high-charge-state oxygen ions, specifically $\rm O^{8^+}$ and $\rm O^{7^+}$, dominate the light impurity population in the SOL near the ICRH antenna limiters. These highly ionized species play a significant role in W sputtering due to their enhanced energy transfer capabilities during ion-surface interactions, driven by their elevated charge states.

\begin{figure}
\includegraphics[width=\linewidth]{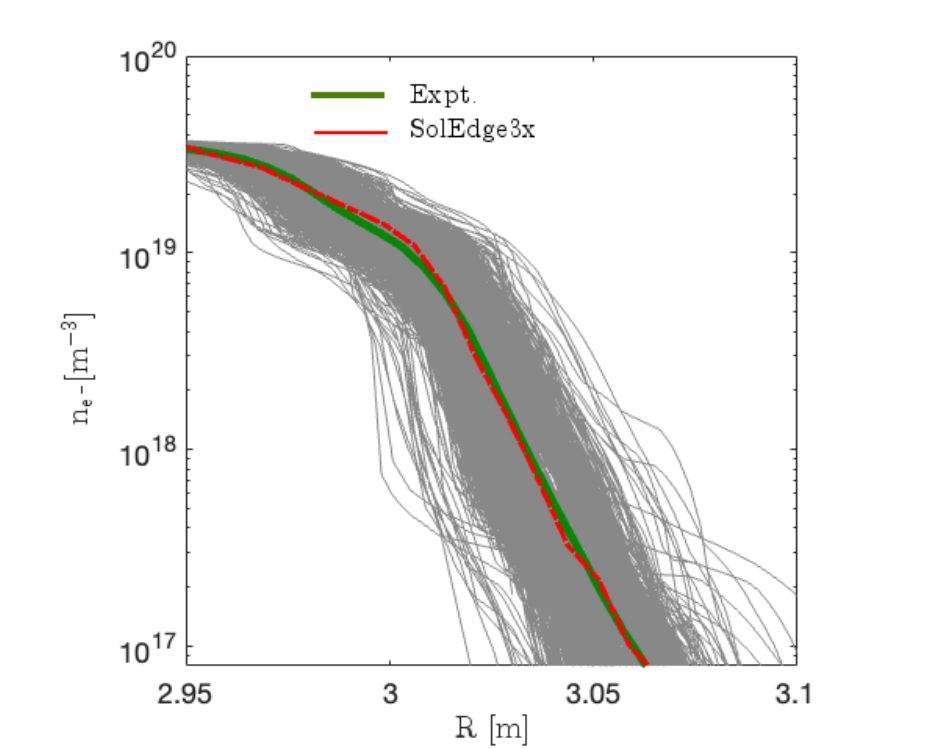}
\caption{Comparison of electron density profile ($\rm n_{e^-}$) at  the OMP from SolEdge3x simulation with $P_{\rm sep}=1.0$~MW (ICRH case) and reflectometry measurements for WEST discharge \#57877. The red curve represents SolEdge3x simulation data, the green curve shows reflectometry data, and the gray traces indicate the experimental time series of $\rm n_{e^-}$.}
\label{fig:5}
\end{figure}

To validate the SolEdge3x simulation data against experimental observations from discharge \#57877, Fig. \ref{fig:5} compares  $\rm n_{e^-}$ at the  OMP, obtained from SolEdge3x with $P_{\rm sep}=1.0$~MW (ICRH case), with reflectometry measurements for WEST discharge \#57877. The red curve shows the simulated $\rm n_{e^-}$ profile, while the green curve represents reflectometry data, and the gray lines depict the time series of experimental $\rm n_{e^-}$ measurements. This comparison demonstrates good qualitative agreement, reinforcing the reliability of the SolEdge3x plasma profiles as background data for STRIPE modeling. As ICRH heating does not significantly impact plasma density, yielding similar $\rm n_{e^-}$ profiles for both ohmic ($P_{\rm sep}=0.4$~MW) and ICRH phases ($P_{\rm sep}=1.0$~MW). However, ICRH increases $\rm T_{e^-}$, though direct $\rm T_{e^-}$ measurements are challenging due to sheath rectification effects, making reflectometry less reliable for $\rm T_{e^-}$ measurement in these conditions. Additionally, no measured $\rm T_{e^-}$ data is available during the ohmic phase ($\rm t<5$~s) for the discharge \#57877.

The fluid simulations provide a steady-state background for STRIPE modeling; thus, the SolEdge3x results will be referred to as "background plasma" in subsequent sections.

\section{Additional Inputs to the STRIPE Framework}
\label{sec:4}

This section details the additional inputs and configurations required for the STRIPE framework to model W erosion at the WEST ICRH antenna limiters accurately.

\subsection{RF Sheath Boundary Conditions in High-Fidelity Wave Solvers} 
\begin{figure}
    \includegraphics[width=\linewidth]{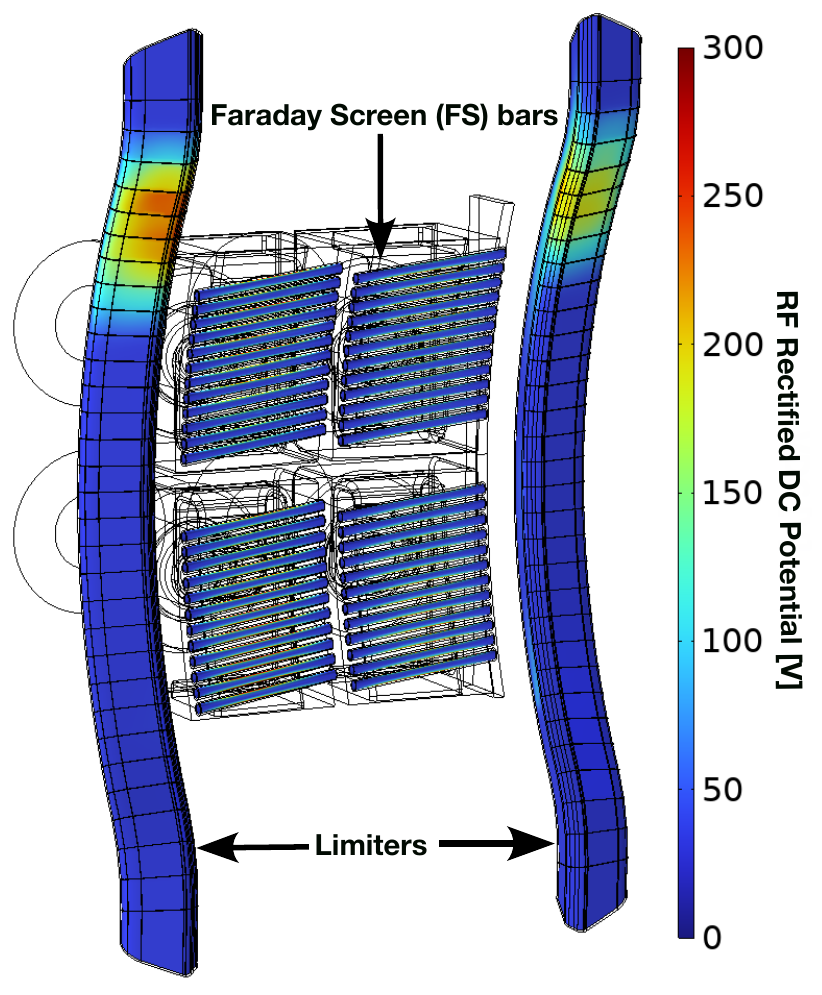}
    \caption{COMSOL-calculated rectified DC sheath voltage (VDC)  on the WEST ICRH antenna structures, including Faraday Screen (FS) bars, showing asymmetry in voltage distribution  \cite{Tierens_2024}. This asymmetry is attributed to variations in electron density, magnetic field orientation, and antenna geometry.}
    \label{fig:6}
\end{figure}

To realistically represent the sheath conditions, the COMSOL model of the WEST antenna calculates RF rectified DC sheath potential  by solving Maxwell’s equations with the cold plasma region surrounding the antenna. Using a sheath-equivalent dielectric layer, the COMSOL model computes rectified voltages on antenna limiters and Faraday Screen (FS) bars, as depicted in Figure~\ref{fig:6}. The COMSOL-calculated RF rectified voltages, shown in Fig.~\ref{fig:6}, display notable asymmetries in the voltage distribution across the antenna structures. These asymmetries are attributed to spatial variations in mainly in  magnetic field orientation, and the antenna’s geometric design.

For the COMSOL modeling, \(n_{e^-}\) data is sourced from reflectometry measurements on WEST discharge \#57877, as shown in Figure~\ref{fig:5}. Due to measurement limitations, \(T_{e^-}\) is set to a fixed 10 eV, representing the approximate average temperature near the antenna. Although this fixed value of $\rm T_{e^-}$ cannot account for rectified sheath voltages during ICRH, it provides a reasonable baseline for potential calculations. The detailed investigation of asymmetries in voltage distribution, including the FS bars in the sheath calculations, is documented in reference \citen{Tierens_2024}. 

\begin{figure}
    \includegraphics[width=\linewidth]{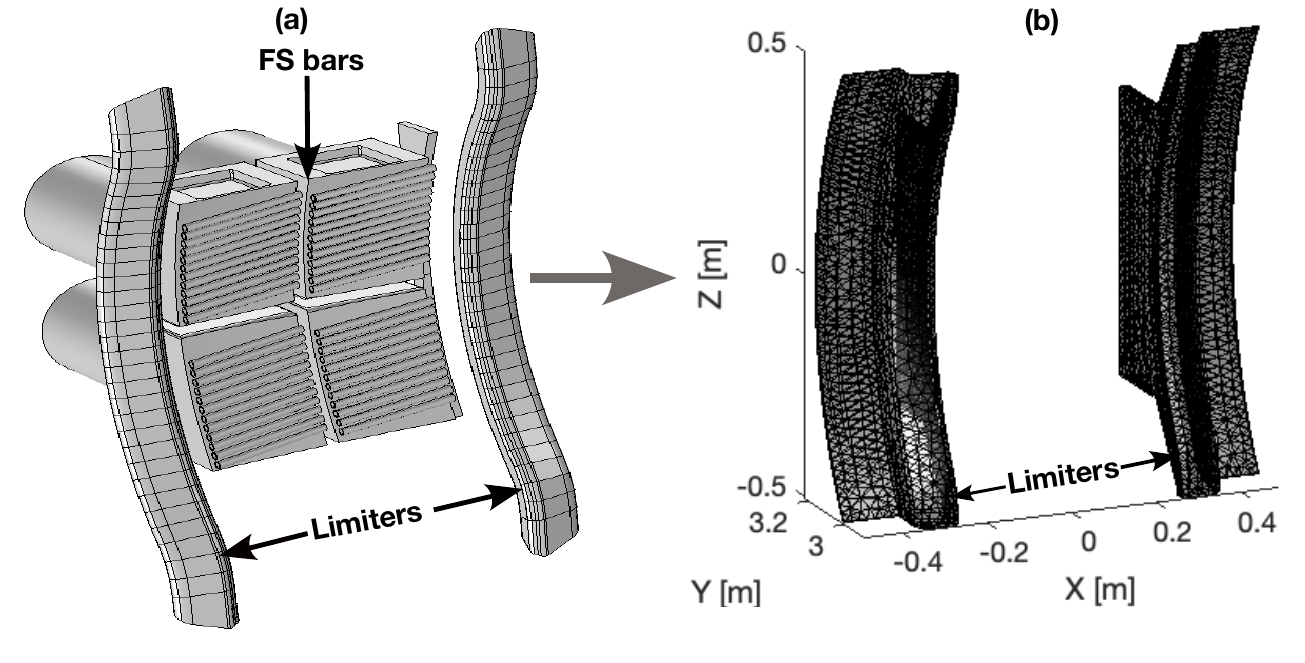}
    \caption{WEST ICRH antenna geometry for STRIPE modeling: (a) Original CAD model of the antenna, and (b) computational geometry with focus on antenna limiters where spectroscopic measurements are available.}
    \label{fig:7}
\end{figure}

\subsection{CAD Defeaturing and Mesh Refinement for ICRH Antenna Structures}
A realistic computational geometry is essential for STRIPE’s finite-element analysis. However, the original CAD model of the WEST ICRH antenna, shown in Figure~\ref{fig:7}a, is defeatured to eliminate small or intricate features that are computationally costly and unnecessary for modeling. This reduction focuses computational resources on regions of interest, such as the antenna limiters, shown in Fig.~\ref{fig:7}b. For this study, the analysis is confined to the antenna limiters where spectroscopic measurements are available, excluding the Faraday Screen (FS). The full antenna geometry will be modeled in future simulations.

\subsection{Tungsten Erosion Yield Calculation}
\begin{figure*}
    \includegraphics[width=\textwidth,height=4cm]{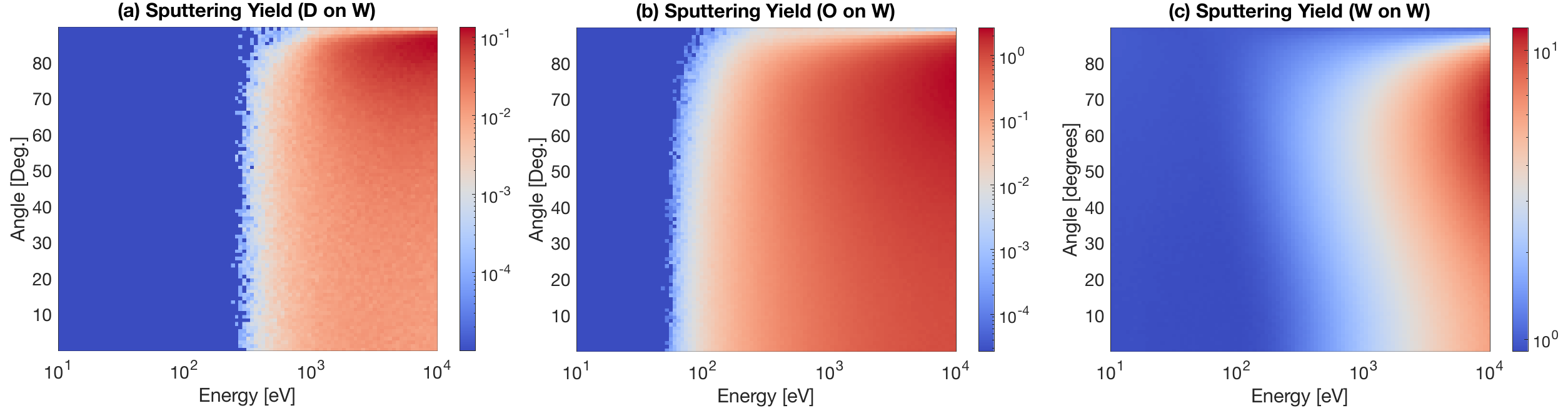}
    \caption{RustBCA-simulated sputtering yield for W  due to impacts from (a) deuterium (D), (b) oxygen (O), and (c) W ions. The yield is shown as a function of incident ion energy and angle, highlighting differences in sputtering behavior across ion types.}
    \label{fig:8}
\end{figure*}

The STRIPE framework integrates RustBCA, a BCA code \cite{RustBCA, Drobny:2023} optimized for simulating ion-material interactions, to calculate the sputtering yield for W under ion impacts from $\rm D^+$, oxygen, and W ions across a range of energies (10 eV to 10,000 eV) and impact angles (0–90°). The sputtering yield results, presented in Figure~\ref{fig:8}, reveal significant dependencies on ion energy and angle. For example, Fig. \ref{fig:8}a shows deuterium’s low sputtering yield across most impact angles with a very high sputtering threshold of $\sim 250$~eV, while Fig. \ref{fig:8}b and \ref{fig:8}c highlight higher yields for oxygen and W ions at specific angles and energies. 

Figure~\ref{fig:8} illustrates general trends in sputtering yield as a function of ion energy and incident angle for various ion species, derived under simplified conditions. These trends do not consider factors such as geometry curvature, magnetic field orientations, sheath effects, or plasma-specific variations.

\subsection{IEADs and Gross W-Erosion Flux ($\rm \Gamma_{\rm gross, W}$)}
To calculate effective sputtering yields for structures like the ICRH antenna, Ion Energy and Angle Distributions (IEADs) must be included to account for the spectrum of ions impacting the surface after passing through the sheath. Furthermore, sputtering on complex geometries, such as antenna limiters, requires a detailed kinetic treatment to incorporate localized plasma variations, magnetic field effects, sheath effects and geometric intricacies.

The particle tracer code-GITR approximates IEADs by launching  10000 ions from the sheath entrance, tracing their trajectories towards the antenna geometry through electric and magnetic fields. Using Maxwellian distributions, these ions (e.g., \(\rm O^{x^+}\) ions) are launched along the magnetic field direction from the sheath entrance and tracked through gyro-motion, drift effects, and collisions. 

Plasma profiles for electrons, deuterium, and oxygen ions, obtained from SolEdge3x simulations as described in Section \ref{sec:3}, are shown in Fig.~\ref{fig:4}. These profiles provide the boundary conditions for the IEADs calculations.

The COMSOL-calculated DC sheath potential (Figure~\ref{fig:6}) also serves as an input for IEADs calculations in GITR. Using this surface potential, the electric field (\(\rm E_n\)) near the antenna surface is modeled by the analytical field expression in GITR:

\begin{equation}
    E_n = V_{\rm sheath} \left[\frac{f_{\rm de}}{2\lambda_{\rm de}} \exp\left(\frac{-r}{2\lambda_{\rm de}}\right) + \frac{1-f_{\rm de}}{\rho_i} \exp\left(\frac{-r}{\rho_i}\right)\right]
    \label{eq:1}
\end{equation}
where \(\rm V_{\rm sheath}\) is the rectified DC sheath potential at the antenna surface; \(\rm f_{de}\) is the fraction of sheath potential difference within the Debye region, and \(\rm r\) represents the normal distance of Debye and Chodura sheath from the surface \cite{Brooks_1990, Brooks_2002, Ding:2016}, $\lambda_{de}$ is the electron Debye length and $\rho_i$ is the ion Larmor radius. The electric field model, coupled with collisions, defines the ion impact energies and angles, affecting the final sputtering yield.

Each surface mesh element, totaling approximately 20,000 for the antenna geometry, undergoes an IEAD calculation to determine the distribution of incident energies (\(\rm E\)) and angles (\(\rm \theta\)). The effective sputtering yield \(\rm Y_{\rm eff}\) for each surface element is computed by integrating over energy and angle:

\begin{equation}
    Y_{\rm eff}(i) = \int_{\theta=0}^{90^\circ} \int_{E=E_{\rm min}}^{E_{\rm max}} Y_i(E, \theta) f_i(E, \theta) \, d\theta \, dE
\end{equation}
where \(i\) indexes each mesh element, and \(\rm E_{\rm min}\) and \(\rm E_{\rm max}\) represent the energy range. This approach ensures that magnetic field angles and antenna curvature, which critically impact erosion, are accounted for.

The W gross erosion flux, \(\rm \Gamma_{gross, W}\) [particles/$\rm m^2$/s], is then calculated by summing the incident ion flux, \(\rm \Gamma_{\rm ions,W}\), weighted by \(\rm Y_{ eff}\) across all surface elements:

\begin{equation}
    \Gamma_{\rm gross,W} = \sum_{i=1}^N Y_{\rm eff}(i) \, \Gamma_{\rm ions}(i)
\end{equation}
\(\rm \Gamma_{\rm gross, W}\) serves as the initial condition in the GITR macroscopic transport simulations, where it initializes particle tracking from antenna structures.

\section{Synthetic Diagnostic Method for Converting W-I Erosion Flux to Photon Emission}
\label{sec:5}

A synthetic diagnostic is developed within STRIPE to convert modeled neutral W (commonly referred as  {\textit {W-I}} in spectroscopy) erosion flux  into brightness (photon emission), $\rm I_\phi$  for direct comparison with experimental spectroscopic data, utilizing the collisional radiative code “ColRadPy” \cite{curt2019}. This diagnostic leverages the inverse photon efficiency or \(\rm S/XB\) coefficients  for W-I impurities, which represents the ionizing flux to photon emission ratio.

\begin{figure}
  \includegraphics[width=\linewidth]{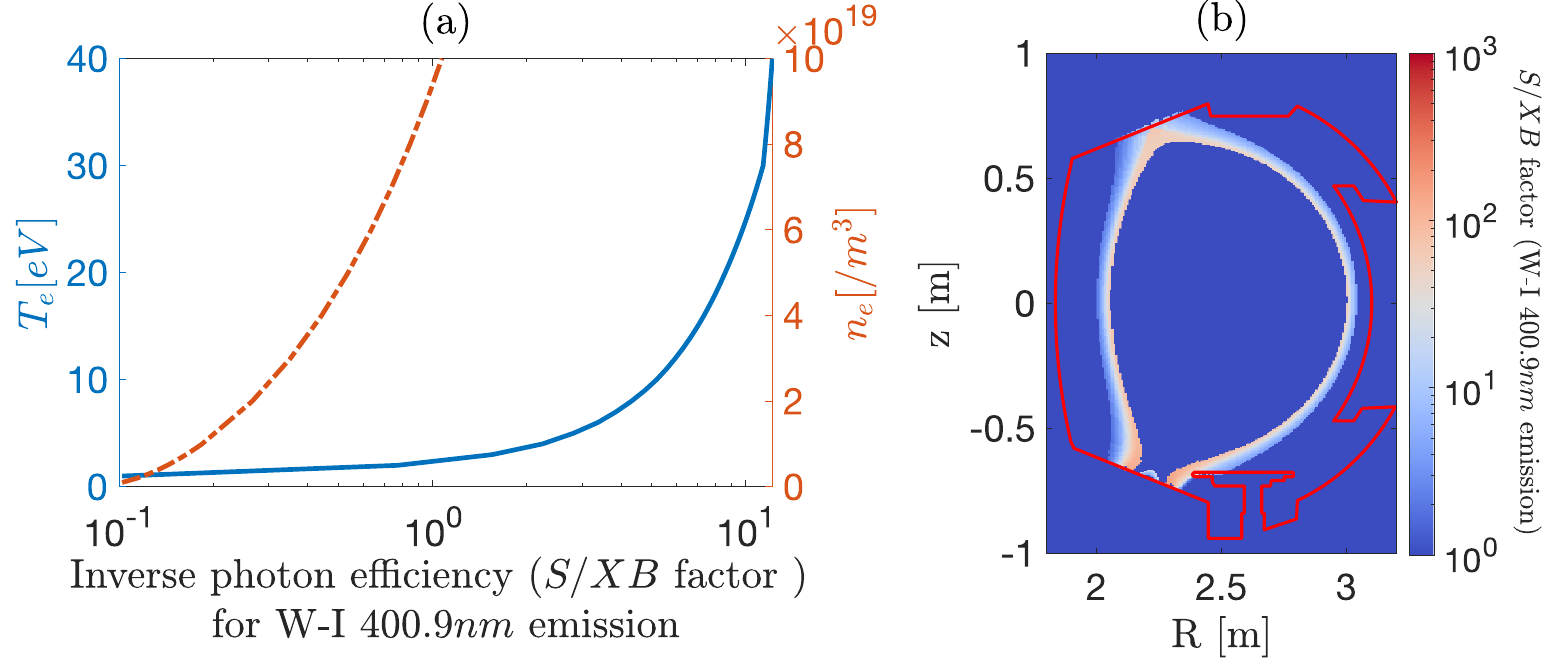}
  \caption{ColRadPy-calculated \(\rm S/XB\) factor for W-I 400.9 nm line emission: (a) \(\rm S/XB\) as a function of \(T_{e^-}\) (blue curve) and \(n_{e^-}\) (red curve); (b) interpolated \(\rm S/XB\) matching SolEdge3x profiles from Fig. \ref{fig:4}.}
  \label{fig:9}
\end{figure}

Figure~\ref{fig:9}a presents the \( \rm S/XB \) factor for the W-I 400.9 nm line emission, calculated using ColRadPy, as a function of \( \rm T_{e^-} \) and \( \rm n_{e^-} \). This factor serves as a basis for linking \( \rm \Gamma_{\rm ions,W} \) to the observable \( \rm I_\phi \). To accurately reflect the plasma conditions in WEST, the \( \rm S/XB \) values are interpolated using the density and temperature profiles from SolEdge3x, depicted in Figure~\ref{fig:4}. The resulting \( \rm S/XB(T_{e^-}, n_{e^-}) \) values, derived from SolEdge3x’s \( \rm T_{e^-} \) and \( \rm n_{e^-} \) for the WEST case, are shown in Figure~\ref{fig:9}b.

The interpolated \(\rm S/XB\) factor is then applied to convert the modeled  \(\rm \Gamma_{gross, W}\), into  \(\rm I_\phi\), using:

\begin{equation}
   \rm  I_\phi = \frac{1}{A} \int \frac{\Gamma_{gross, W}}{4\pi \rm S/XB} \, dA
\end{equation}

where \(A\) represents the optical viewing area with radius set to 90 mm as per previous diagnostic measurements \cite{Meyer_2018}.

This synthetic diagnostic validates STRIPE’s model predictions by comparing calculated $\rm I_\phi$ with spectroscopic measurements from WEST discharge \#57877, benchmarking the framework’s performance in capturing RF-induced sheath effects and ion impacts on W erosion.

\section{Results: W-Erosion at the WEST ICRH Antenna Limiters}
\label{sec:6}
This section presents a comprehensive analysis of W erosion at the WEST ICRH antenna limiters, estimated using the STRIPE framework. We provide detailed erosion calculations for both thermal and RF sheath conditions during the ICRH phase, with validation against experimental data. Additionally, we analyze \( \rm \Gamma_{{gross, W}} \) due to oxygen impinging on W limiters, \( \rm \Gamma_{{gross, W}}(O^{x^+} \rightarrow W) \) and corresponding effective sputtering yield $\rm  Y_{{eff}}(O^{x^+} \rightarrow W)$, in the ohmic phase, offering insights into W erosion behavior under varied plasma conditions and thermal sheath scenarios.

\subsection{Thermal Sheath Case: ICRH Phase Erosion Results}
In this subsection, we analyze $\rm  \Gamma_{{gross, W}}(O^{x^+} \rightarrow W)$ obtained from the STRIPE framework under thermal sheath conditions. The simulations incorporate the plasma parameters and specific inputs outlined in Sections \ref{sec:3} and \ref{sec:4}. The background plasma profiles for the ICRH phase are chosen from the SolEdge3x simulations with $\rm P_{sep}=1$~MW.

The thermal sheath voltage on the ICRH antenna structures is calculated \cite{stangeby:2000} as:
\begin{equation}
  \rm   V_{{sheath}}({thermal}) = k_{{sheath}} \times T_{e,{surf}}
    \label{eq:5}
\end{equation}
where the sheath factor, $\rm( k_{{sheath}} = \rm {abs}[0.5 \times \log(2\pi m_e/m_i)][1 + T_{i,{surf}}/T_{e,{surf}}]$, depends on the electron and ion masses $m_e$ and $m_i $, and the ion and electron temperatures $\rm  T_{i,{surf}}$ and $\rm T_{e,{surf}}$ respectively at the antenna surfaces. For surface temperatures $\rm T_{{surf}} \sim 10$~{eV}, this yields a peak thermal sheath voltage of approximately 30 eV.

\begin{figure*}
    \includegraphics[scale=0.4]{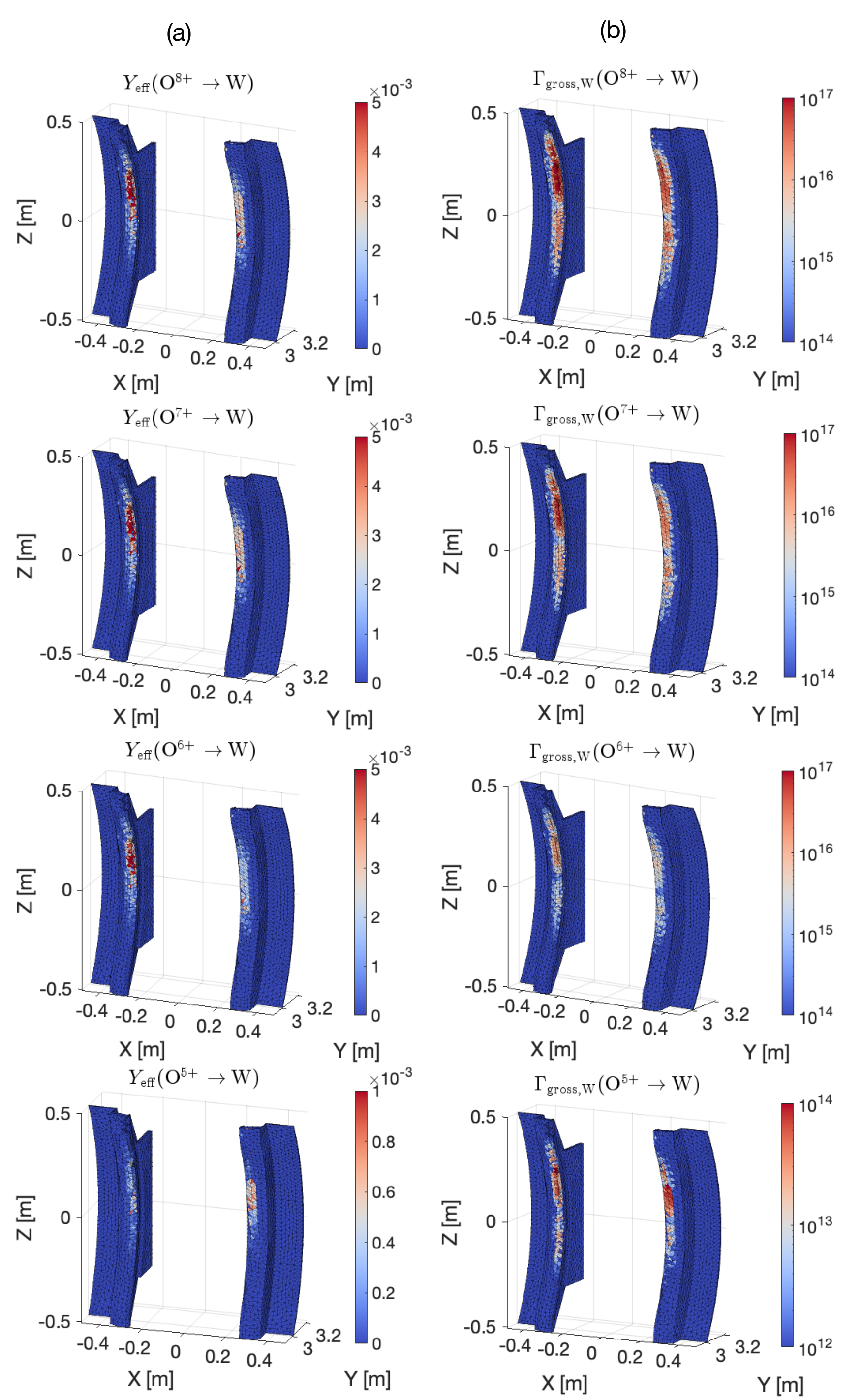}
    \caption{3D maps of (a) effective sputtering yield, $\rm  Y_{{eff}}(O^{x^+} \rightarrow W)$, and (b) gross W erosion flux, $\rm  \Gamma_{{gross, W}}(O^{x^+} \rightarrow W)$, for high-charge-state oxygen ions ($\rm O^{6^+}$ and above) impinging on the WEST ICRH antenna limiters under thermal sheath conditions. These maps demonstrate the spatial distribution of sputtering yield and erosion flux, revealing a pronounced erosion concentration on the front-facing surfaces of the limiters where ions strike perpendicularly. The results highlight the substantial contribution of higher oxygen charge states-$\rm  O^{6^+}$ and $\rm  O^{8^+}$  to W sputtering, with $\rm  Y_{{eff}}(O^{8^+} \rightarrow W)$  yielding fluxes orders of magnitude higher than those of lower charge states.}
    \label{fig:10}
\end{figure*}

\begin{figure*}
    \includegraphics[scale=0.4]{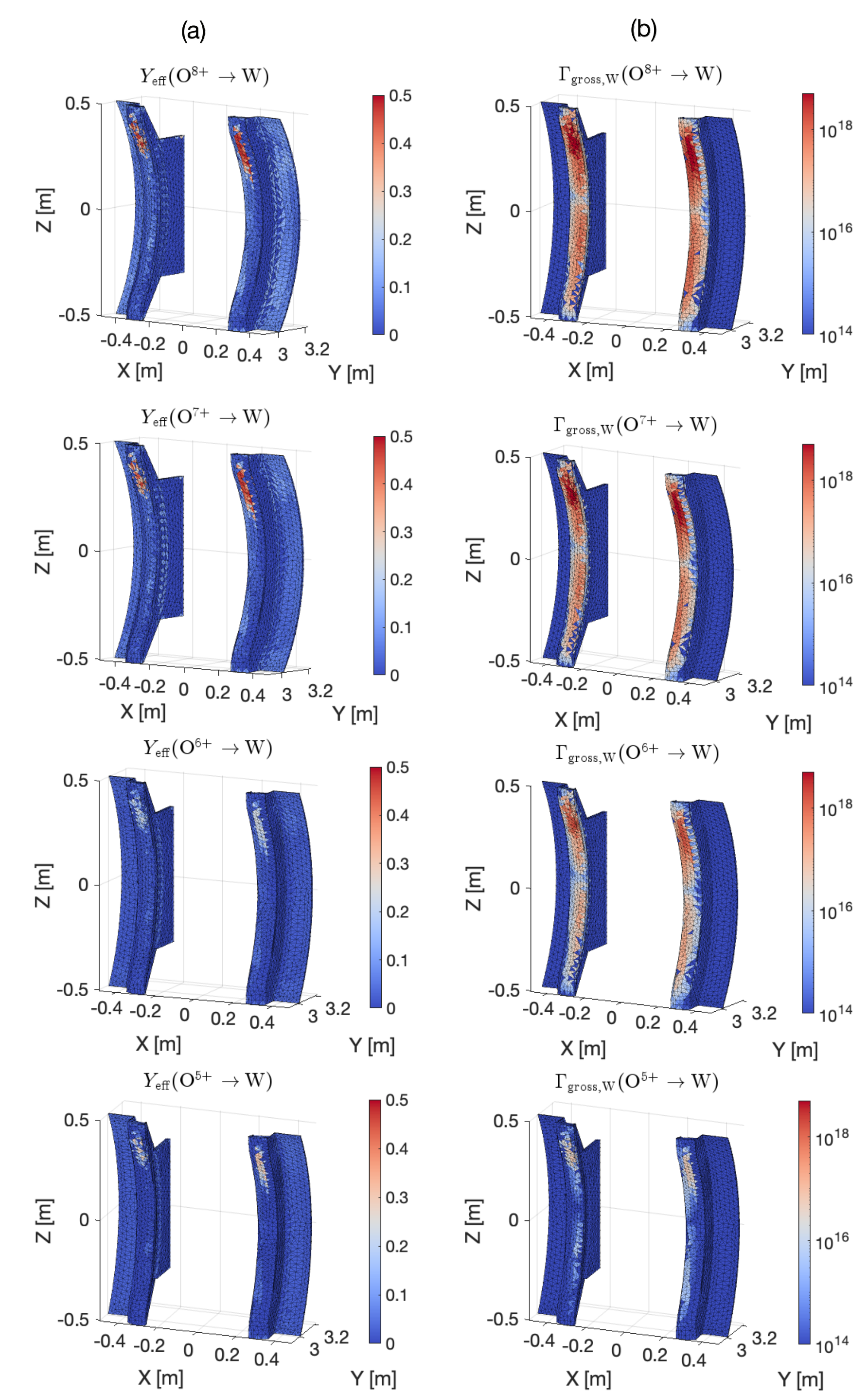}
    \caption{3D maps of (a) $\rm  Y_{{eff}}(O^{x^+} \rightarrow W)$, and (b) $\rm  \Gamma_{{gross, W}}(O^{x^+} \rightarrow W)$, for high-charge-state oxygen ions ($\rm  O^{5^+}$ and above) impacting the WEST ICRH antenna limiters under RF sheath conditions. Compared to thermal sheath conditions, the RF sheath environment yields significantly higher sputtering and erosion rates, with RF sheath rectified potentials calculated by COMSOL reaching approximately 300 eV. This enhancement primarily stems from the increased ion energies of high-charge oxygen ions such as $\rm  O^{6^+}$ and $\rm  O^{8^+}$, leading to tenfold increases in sputtering and erosion fluxes. The RF sheath conditions also introduce marked spatial asymmetry, with erosion hotspots concentrated on the upper regions of the limiters.}
    \label{fig:11}
\end{figure*}

Figure \ref{fig:10} illustrates 3D maps of (a) $\rm  Y_{{eff}}(O^{x^+} \rightarrow W)$, and (b) $\rm  \Gamma_{{gross, W}}(O^{x^+} \rightarrow W)$, for higher charge states of oxygen ($\rm  O^{5^+}$ and above) impinging on the WEST W antenna limiters. The results reveal clear distinctions in erosion contributions among oxygen charge states. For example, \(\rm  Y_{{eff}}(O^{5^+} \rightarrow W) \) is approximately five times lower than $\rm  Y_{{eff}}(O^{8^+} \rightarrow W)$, resulting in $\rm  \Gamma_{{gross, W}}(O^{x^+} \rightarrow W)$ for \(\rm  O^{5^+} \) that are three orders of magnitude smaller than those from $\rm  O^{8^+}$. Consequently, higher charge states of oxygen, specifically \(\rm O^{6^+} \) and above, contribute substantially to W sputtering under thermal sheath conditions, while lower-charge states are less significant.

RustBCA calculations (Fig. \ref{fig:8}) support these findings by indicating that the threshold energy for W sputtering by $\rm D^+$ ions is approximately 250 eV—significantly higher than $\rm V_{{sheath}}({thermal}) $. This suggests that background $\rm D^+$ ions do not contribute meaningfully to sputtering in the thermal sheath cases. Thus, background $\rm D^+$ ions are effectively ruled out as sputtering species, as their sputtering energy threshold also exceeds the thermal sheath potential, $\rm V_{sheath}(thermal)$.

Spatially, erosion is predominantly concentrated on the front-facing surfaces of the ICRH antenna limiters, as shown in Figure \ref{fig:10}. This distribution implies that ions impinging perpendicularly onto these surfaces are more effective in generating sputtering events. Furthermore, the gross erosion profiles exhibit asymmetry along the vertical axis around the OMP. This lack of symmetry is attributable to variations in plasma density, magnetic field profiles, and localized sheath effects, which modulate erosion rates across different areas of the antenna. Moreover, a dip in the gross erosion flux at the OMP  is also observed where the magnetic flux lines are tangential to the limiter surface. 

\subsection{RF Sheath Case: ICRH Phase Erosion Results}
This section provides a detailed analysis of gross erosion results under RF sheath conditions, focusing on the influence of RF rectified sheath potentials calculated using COMSOL. The simulations specifically use the rectified sheath potential on the WEST ICRH antenna, as detailed in reference \citen{Tierens_2024}. According to COMSOL simulations, the rectified sheath potential reaches a peak voltage of approximately 300 eV, nearly ten times higher than the potential observed under thermal conditions. This substantial increase demonstrates the significant impact of RF effects on sheath dynamics and material interaction.

Figure \ref{fig:11} presents 3D maps of two key parameters: (a) $\rm Y_{{eff}}(O^{x^+} \rightarrow W)$, and (b) $\rm \Gamma_{{gross, W}}(O^{x^+} \rightarrow W)$, under RF sheath conditions. These maps focus on oxygen ions with charge states $\rm O^{5^+}$ and higher, providing valuable insights into the erosion behavior influenced by RF-driven plasma conditions. The data show consistent trends with the thermal sheath scenario, where high-charge-state oxygen ions, such as $\rm O^{8^+}$, contribute significantly more to W erosion compared to lower-charge states.

Quantitative analysis reveals that $\rm  Y_{{eff}}(O^{x^+} \rightarrow W)$ from $\rm O^{5^+}$ ions is three times lower than that of $\rm O^{8^+}$. This difference results in $\rm \Gamma_{{gross, W}}(O^{x^+} \rightarrow W)$ for $\rm O^{5^+}$ ions that are two orders of magnitude smaller than those caused by $\rm O^{8^+}$. 

Background $\rm D^+$ ions, despite their presence, have negligible contributions to sputtering in the RF sheath environment. Their sputtering energy threshold barely reaches the peak value of the RF sheath potential, $\rm V_{sheath}(RF)$, limiting their impact. This observation confirms the dominant role of oxygen ions, particularly those with higher charge states, in gross erosion under RF sheath conditions.

In the RF sheath environment, erosion patterns become highly asymmetric, with most erosion occurring on the upper regions of the limiters. This asymmetry arises from the rectified RF sheath potentials calculated in the COMSOL simulations, which show peak values concentrated in the upper parts of the limiters. As also discussed later in Section \ref{subssec:72}, main contributing factors include variations in plasma density, magnetic field configurations, and localized sheath effects across the antenna structure. These factors collectively result in localized ion acceleration to higher energies in specific regions. Conversely, regions near the outboard midplane (OMP) ($\rm z=0~m$), where magnetic flux lines run tangentially to the limiter surface, show a reduction in gross erosion flux.

A comparison of Figures \ref{fig:10} and \ref{fig:11} highlights that RF sheath conditions lead to a tenfold increase in $\rm \Gamma_{{gross, W}}(O^{x^+} \rightarrow W)$ compared to thermal sheath conditions. This enhancement is attributed to higher ion energies, particularly from high-charge oxygen ions such as \( \rm O^{6^+} \) and \( \rm O^{8^+} \). These results underscore the pivotal role of RF sheath potentials in accelerating erosion rates at antenna limiters and emphasize the need for detailed consideration of RF effects in plasma-material interaction models.

\subsection{Integrated Tungsten Erosion Flux at the ICRH Antenna Limiters}

The area integrated (referred simply as {\textit{integrated}} henceforth) W gross erosion flux, $\rm G_{{gross, W}} (O^{x^+}\rightarrow W)$, at the WEST ICRH antenna limiters is analyzed for RF sheath and thermal sheath conditions during the ICRH phase, as well as for thermal-sheath-only conditions during the ohmic phase. Table~\ref{tab:table1} highlights the contributions of different ion species to $\rm G_{{gross, W}} (O^{x^+}\rightarrow W)$ and the total flux for each case.

\begin{table*}
\caption{\label{tab:table1} Integrated W erosion flux, $G$ [particles/s] calculated using STRIPE modeling at the WEST ICRH antenna limiters.}

\begin{tabularx}{\textwidth}{XX}

\begin{tabular}{lcccccc}
\hline \hline
Sheath Condition & $\rm O^{8^+}\rightarrow W$ & $\rm O^{7^+}\rightarrow W$ & $\rm O^{6^+}\rightarrow W$ & $\rm O^{5^+}\rightarrow W$ & $\rm D^+\rightarrow W$ & Total Flux \\
\hline
RF  (ICRH) & $2.263 \times 10^{17}$ & $1.681 \times 10^{17}$ & $4.02 \times 10^{16}$ & $1.109 \times 10^{15}$ & $9.673 \times 10^{11}$ & $4.348 \times 10^{17}$ \\
Thermal (ICRH) & $1.928 \times 10^{16}$ & $1.101 \times 10^{16}$ & $2.579 \times 10^{15}$ & $8.95 \times 10^{14}$ & 0 & $3.285 \times 10^{16}$ \\
Thermal  (ohmic) & $8.05 \times 10^{15}$ & $4.86 \times 10^{15}$ & $6.45 \times 10^{14}$ & $2.66 \times 10^{14}$ & $0$ & $1.385 \times 10^{16}$ \\
\hline
\end{tabular}
\end{tabularx}
\end{table*}

The results reveal that under RF sheath conditions during the ICRH phase, high-charge-state oxygen ions (\(\rm O^{8^+}\)) dominate W erosion, contributing approximately \(2.26 \times 10^{17}\) particles/s. This is more than \(200\)-fold higher than the contribution from \(\rm O^{5^+}\) (\(1.11 \times 10^{15}\) particles/s) and approximately \(2.3 \times 10^5\)-times higher than $\rm D^+$ ions , which contribute negligibly at \(9.67 \times 10^{11}\) particles/s. The total erosion flux under RF sheath conditions is the highest among all cases, reaching \(4.35 \times 10^{17}\) particles/s.

For thermal sheath conditions during the ICRH phase, the erosion contribution from \(\rm O^{8^+}\) ions reduces to \(1.93 \times 10^{16}\) particles/s, about \(8.5\%\) of their contribution under RF sheath conditions. The contribution from \(\rm O^{5^+}\) ions in this phase is \(8.95 \times 10^{14}\) particles/s, approximately \(22\)-times lower than that of \(\rm O^{8^+}\). $\rm D^+$ ions do not contribute to erosion under thermal sheath conditions due to insufficient energy for sputtering. The total flux in this scenario is significantly lower at \(3.29 \times 10^{16}\) particles/s, about \(7.6\%\) of the flux under RF sheath conditions.

During the ohmic phase, characterized solely by thermal sheath conditions, the erosion flux reduces further for all ion species. \(\rm O^{8^+}\) ions contribute \(8.05 \times 10^{15}\) particles/s, approximately \(42\%\) of their contribution during the thermal sheath in the ICRH phase and about \(3.5\%\) of their flux under RF sheath conditions. \(\rm O^{5^+}\) ions contribute \(2.66 \times 10^{14}\) particles/s, which is \(30\)-times lower than \(\rm O^{8^+}\). The total flux in this case is the lowest, at \(1.39 \times 10^{16}\) particles/s, only \(3.2\%\) of the RF sheath case total flux.

These comparisons emphasize the dominant role of high-charge-state oxygen ions (\(\rm O^{8^+}\)) in driving W erosion, particularly under RF sheath conditions, where their contributions far exceed those of \(\rm O^{5^+}\) or $\rm D^+$ ions. The significant differences in total erosion fluxes between RF and thermal sheath conditions underscore the critical impact of RF-induced sheath potentials on W erosion dynamics.

\subsection{Validation with Experimental Observations}
To validate the $\rm  \Gamma_{{gross, W}}(O^{x^+} \rightarrow W)$, STRIPE’s model predictions (assuming a plasma composition of 99\% $\rm D^+$ and 1\% oxygen) are compared with spectroscopic data from WEST discharge \#57877 during both ICRH and ohmic phases. Using $\rm S/XB$ coefficients, STRIPE’s synthetic diagnostics converts the modeled $\rm  \Gamma_{{gross, W}}(O^{x^+} \rightarrow W)$ into $\rm I_\phi$ (Section \ref{sec:5}), allowing for direct comparison with observed W-I line emissions. The spectroscopic measurements were taken along specific vertical sightlines at z = 0.4m, 0.3473m, 0.1689m, 0.0749m, 0.0106m, -0.0851m, -0.2m, -0.3042m and  -0.4m at the ICRH antenna limiters, as shown by the green lines in Fig.~\ref{fig:4}.

\subsubsection{ICRH Phase Validation:}

\begin{figure}
    \includegraphics[width=\linewidth]{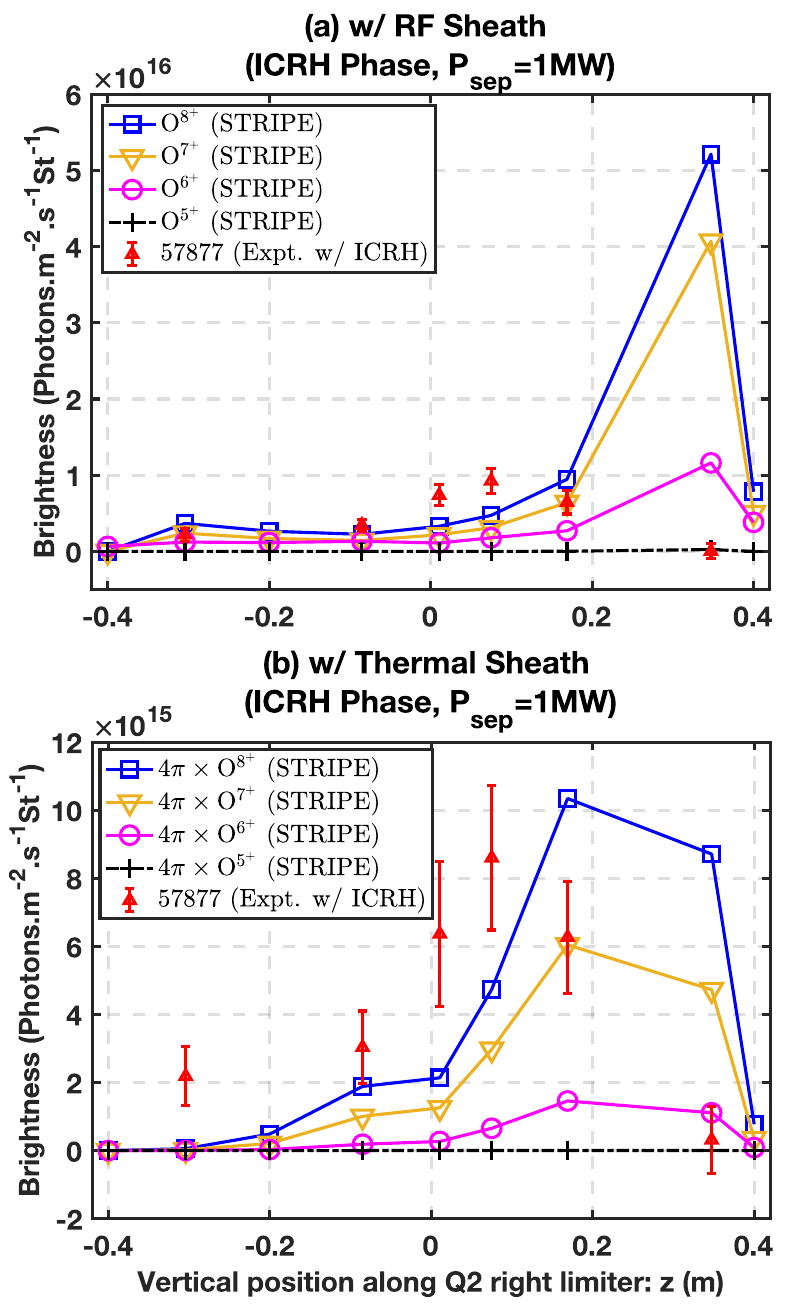}
    \caption{Comparison of modeled and experimental W-I 400.9 nm emissions at the WEST ICRH antenna limiter during the ICRH phase of discharge \#57877. (a) RF sheath case shows enhanced $\rm I_\phi$, with a notable peak at z =0.3473 matching COMSOL-predicted hotspots. (b) Thermal sheath case shows $\rm I_\phi$ an order of magnitude lower, with slight spatial offsets attributed to local plasma variations. These results validate STRIPE's accuracy in capturing sheath-induced W sputtering dynamics.}
    \label{fig:12}
\end{figure}

Fig. \ref{fig:12}a shows $\rm I_\phi$  obtained for the RF sheath case during the ICRH phase Notably, STRIPE's simulated $\rm I_\phi$ for various oxygen charge states in the RF sheath case align closely with experimental observations, except at $z=0.3473$~m where the model predicts a peak $\rm  \Gamma_{{gross, W}}(O^{x^+} \rightarrow W)$. This peak arises from the rectified potential distribution simulated by COMSOL, which indicates a prominent hot spot at the same location on the antenna (Fig. \ref{fig:6}). These findings demonstrate the critical contribution of RF sheath rectification to W erosion and impurity generation. Moreover, higher charge states of oxygen (\(\rm O^{6^+} \) and above) are identified as dominant sputtering species under these conditions.

Figure \ref{fig:12}b shows $\rm I_\phi$ for the thermal sheath case during the ICRH phase, mapped for different oxygen charge states. Compared to the RF sheath case, W-I line emissions under thermal sheath conditions are about ten times lower. While the model with thermal-sheath-only agrees well with  experimental observations quantitatively, a slight spatial shift is evident in $\rm I_\phi$ peak location. 

\subsubsection{Ohmic Phase Validation:}
\begin{figure}
    \includegraphics[width=\linewidth]{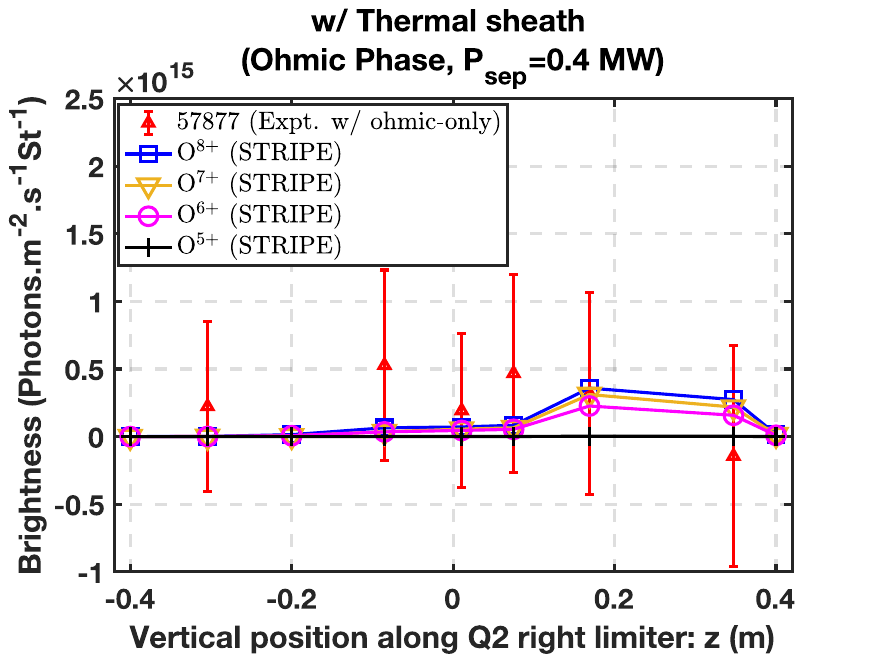}
    \caption{Comparison of modeled and experimental W-I 400.9 nm emissions at the WEST ICRH antenna limiter during the ohmic phase of discharge \#57877. The thermal sheath model shows quantitative agreement with spectroscopic data with  a slight spatial shift  in the flux peak location. }
    \label{fig:13}
\end{figure}

Further validation of the STRIPE framework was performed using spectroscopic data from WEST discharge \#57877 during the ohmic phase. Background plasma profiles for this phase were obtained from SolEdge3x simulations with  $ \rm P_{sep}= 0.4$~MW. The calculated $\rm  \Gamma_{{gross, W}}(O^{x^+} \rightarrow W)$ was converted into $\rm I_\phi$ using STRIPE’s synthetic diagnostics (see Section \ref{sec:5}). Fig. \ref{fig:13} compares modeled and experimental W-I line emissions along vertical sightlines, demonstrating a quantitative  agreement within experimental error bars. However, a slight spatial shift is also observed in the $I_\phi$ peak location. This discrepancy may stem from  variations in plasma parameters, such as density or magnetic field profiles, or localized sheath effects influencing ion impact energies. Such factors could result in localized erosion variations not fully captured by the model and are discussed later in Section \ref{subssec:72}.

During the ohmic phase, which is characterized solely by thermal sheath conditions, $\rm \Gamma_{{gross, W}}$ is significantly lower than during the ICRH phase. The reduced sheath potential in this phase resulted in lower ion impact energies and minimal W erosion. 

The discrepancies between modeling and experiments, observed in Figures~\ref{fig:12} and \ref{fig:13}, are discussed in detail in Sections~\ref{subssec:72}-\ref{subssec:74}.

\section{Discussion}
\label{sec:7}

The results presented in this study provide a comprehensive understanding of W erosion at the ICRH antenna limiters in the WEST tokamak, emphasizing the distinct impacts of thermal and RF sheath conditions during the ICRH phase. Through detailed simulations validated against experimental observations, this work underscores the critical role of RF-induced sheaths in amplifying erosion rates at plasma-facing components. This discussion highlights the implications of these findings for PMI in fusion devices, potential improvements to the STRIPE framework, and key considerations for future fusion reactor designs as discussed in Section \ref{subssec:76}.

\subsection{The Role of RF Sheaths in Enhancing W Erosion}

A key insight from this study is the tenfold increase in $\rm \Gamma_{gross, W}(O^{x^+} \rightarrow W)$ observed in RF sheath conditions relative to thermal sheaths during the ICRH phase. The comparison between these two cases, as illustrated in Fig. \ref{fig:12}a-b, reveals that RF sheath potentials significantly amplify ion energies by accelerating high-charge-state ions, particularly oxygen ions (\( \rm O^{6^+}\) and higher), to impact energies well beyond the thermal sheath potential range. This enhancement in ion energy directly translates to elevated sputtering yields and, consequently, higher W erosion rates\cite{Klepper2022, Hakola:2021}. The STRIPE modeling suggests a more or less  linear relation between $\rm \Gamma_{gross, W}(O^{x^+} \rightarrow W)$ and sheath voltage at the antenna limiters. 

The considerable difference in W erosion between thermal and RF sheath cases underscores the impact of ICRH on PMI processes. With RF sheaths, rectified potentials create conditions that favor higher-energy impacts from impurity ions, thereby intensifying sputtering. This effect has significant practical implications for fusion reactor operations, particularly in high-power ICRH scenarios where RF-induced sheaths can accelerate material degradation of PFCs.

\subsection{Spatial Asymmetry in Erosion Profiles}
\label{subssec:72}

The spatial distribution of W erosion, particularly the pronounced asymmetry observed under RF sheath conditions, underscores the significant impact of RF sheaths and the geometric characteristics of the WEST ICRH antenna on erosion patterns. As illustrated in the 3D erosion maps, erosion hotspots are predominantly concentrated in the upper regions of the antenna limiters during RF sheath conditions. This asymmetry arises from a combination of factors, including variations in plasma density, magnetic field orientation, localized sheath effects, and the complex RF field distribution modeled using COMSOL \cite{Easley:2024}.

Reference \citen{Tierens_2024} indicates that while plasma density and magnetic equilibrium  variations influence the magnitude of sheath voltage, they do not alter the spatial location of erosion hotspots. Future investigations, such as magnetic field angle scans, are planned to provide further insights into these effects.

In the current study, plasma profiles are derived from 2D SolEdge3x simulations, which assume axisymmetric walls and plasma conditions. These profiles are extrapolated onto the 3D geometries of the WEST ICRH antenna, which in reality exhibits significant geometric asymmetry along the vertical coordinate. This inherent asymmetry in the antenna geometry likely affects the observed erosion patterns and may not be fully captured by the current modeling approach. To better understand the spatial asymmetry in erosion profiles, future work will explore the use of SolEdge3x simulations performed directly in 3D. 

The current numerical scheme in the SolEdge3x code does not yet account for biased voltages on antenna walls or other plasma-facing components (PFCs) arising from sheath formation at the plasma-material interface. RF antenna sheaths can bias the edge plasma potential, resulting in the formation of steady-state convective cells in the scrape-off layer (SOL). This leads to RF-induced DC E×B particle convection transverse to the magnetic field lines \cite{Ippolito:1993, Kubic:2012, Tamain:2017}, which can significantly influence particle fluxes and local plasma conditions near the ICRH  antenna. To enhance the accuracy of plasma-antenna interaction modeling, future simulations will include these effects, capturing the role of E×B drifts in SOL plasma transport dynamics.

The localized erosion hotspots identified in this study have significant implications for PFCs. Uneven material wear from geometric asymmetries and RF sheath effects can cause premature failure and increased maintenance. Accurate modeling of rectified RF sheath potentials and detailed spatial simulations of complex 3D geometries, like the WEST ICRH antenna, are essential for enhancing PFC durability and optimizing PMI strategies in future fusion devices.

\subsection{Validation with Experimental Observations}
\label{subssec:73}
Validation of STRIPE against experimental observations from WEST discharge \#57877 demonstrates the framework's reliability in interpreting and predicting W erosion dynamics under both thermal and RF sheath conditions.

\textbf{ICRH Phase:} During the RF sheath phase, STRIPE predictions closely align with measured W-I 400.9 nm emissions, except for a localized peak at \(z=0.3473\)~m, attributed to rectified sheath potentials simulated by COMSOL. This agreement underscores the dominant role of high-charge oxygen ions  in RF-enhanced erosion. The thermal sheath case predicts $I_\phi$ an order of magnitude lower, consistent with experimental trends, although a slight spatial offset in the flux peak is observed. This offset may result from unaccounted plasma variations or localized effects.

\textbf{Ohmic Phase:} Under thermal sheath-dominated conditions, STRIPE quantively reproduces the significantly lower erosion rates observed experimentally, with results falling well within error bars. The reduced ion energies during this phase result in minimal erosion, further validating STRIPE's reliability in scenarios governed by thermal sheath interactions.

To achieve reasonable agreement between STRIPE predictions and experimental data from WEST, the present study assumes a plasma composition of 99\% $\rm D^+$ and 1\% oxygen. Oxygen serves as a proxy for all light impurities present during WEST discharge \#57877. However, recent experiments on WEST frequently employ boron powder injections for wall conditioning \cite{Bodner:2022, Afonin:2024}. Future STRIPE modeling efforts will investigate the contribution of boron and other light impurities to W erosion at the ICRH antenna, providing a more comprehensive understanding of impurity-driven erosion dynamics.

\subsection{Model Sensitivity and Reliability}
\label{subssec:74}

The reliability of STRIPE calculations is influenced by the sensitivity of the model to various input parameters and assumptions. Plasma background profiles, sheath potential estimates, and ion energy-angle distributions significantly impact erosion predictions. Even minor deviations in plasma density, magnetic field strength, or sheath properties can shift erosion profiles and intensities. 

Discrepancies observed in erosion peak locations, particularly under RF sheath conditions, suggest areas for improvement. These deviations may arise from:

\begin{itemize}
    \item Variations in RF sheath potentials simulated by COMSOL.
    \item Uncertainties in plasma parameters or localized sheath effects influencing ion impact energies.
\end{itemize}

Additionally, STRIPE’s predictions are sensitive to light impurity concentrations. For instance, the current study assumes a 1\% oxygen concentration, but other light impurities, such as boron, and their concentrations may also impact the results.

To address these sensitivities, incorporating higher-fidelity tools and self-consistent models—such as fully kinetic plasma solvers or coupled turbulence-sheath simulations—could improve input accuracy. However, these tools often come with substantial computational costs, presenting challenges for large-scale simulations. Balancing model accuracy with computational feasibility will be essential as simulation demands increase for next-generation fusion reactors.

Addressing these variations through enhanced plasma and sheath modeling will further improve STRIPE’s predictive accuracy, ensuring its reliability in simulating erosion dynamics across a broad range of fusion scenarios.

\subsection{Enhanced Modeling Capabilities and Future Directions for STRIPE Framework Development}
\label{subssec:75}
The STRIPE framework currently integrates GITR, enabling detailed calculations of gross erosion, net erosion, and re-deposition profiles, alongside global transport simulations of sputtered impurities. These capabilities are essential for simulating W self-sputtering effects, allowing STRIPE to produce comprehensive profiles of material loss, re-deposition, and impurity migration over time. This integrated approach provides a thorough assessment of net erosion impacts on PFCs and offers key insights into impurity redistribution in the whole device, particularly during high-power ICRH operations. 

Within the current framework, GITR supports the analysis of net erosion and re-deposition on PFCs including ICRH antenna structures, capturing the dynamics of material redistribution due to  erosion and self-sputtering on the surface. Additionally, GITR’s advanced global transport capabilities enable efficient and detailed simulation of the migration of sputtered impurities within the plasma environment on global scale.  By modeling impurity migration, GITR can analyze how erosion, transport, and re-deposition patterns affect plasma-facing components (PFCs). These insights are essential for optimizing PFC design, mitigating impurity-induced plasma contamination, and improving the overall performance of fusion devices. Recent studies on linear devices, such as Proto-MPEX \cite{Beers1, Beers2, Rapp:2024} and PISCES-RF \cite{Dhamale:2024}, have demonstrated the utility of these capabilities in analyzing impurity dynamics and material interaction processes.

Looking ahead, we plan to integrate GITRm, an advanced meshing variant of GITR, which incorporates adaptive meshing capabilities as detailed in reference \citen{nath:2023}. GITRm’s adaptive meshing will enable STRIPE to simulate the complete and intricate antenna geometry, including FS bars and other complex structures that were simplified in this study. This enhanced capability will allow for detailed modeling of the 3D interactions between RF-induced sheath potentials and magnetic fields across the full antenna assembly, capturing the localized effects of geometry on erosion and re-deposition. The inclusion of GITRm is anticipated to provide a more precise analysis of erosion, re-deposition, and impurity transport, essential for evaluating PFC durability and effectiveness under intense ICRH conditions. 

Additionally, incorporating temperature-dependent sputtering yields, crystallographic effects \cite{Samolyuk:2023}, and self-consistent turbulence modeling within the SolEdge3x simulations \cite{Bufferand:2024} could provide a more nuanced understanding of PMI in fusion reactors. Refining background plasma profiles with advanced turbulence models could further improve the accuracy of simulated ion fluxes and sheath potentials, particularly in high-performance fusion reactors where turbulence-driven transport affects sheath dynamics.

To streamline code coupling and improve data exchange within the STRIPE framework, the integration of an Integrated Plasma Simulation (IPS) framework \cite{Hoenen:2015} is planned. This enhancement aims to extend STRIPE’s applicability across a broader range of fusion devices and operational scenarios. 

\subsection{Implications for Future Fusion Experiment Design}
\label{subssec:76}
The findings of this study have significant implications for the design and operation of next-generation fusion devices, such as ITER, SPARC, and MPEX, where high-power ICRH systems are expected to play a critical role in plasma heating. The elevated erosion rates observed under RF sheath conditions indicate that PFCs exposed to RF-induced sheaths will undergo accelerated material degradation, potentially shortening their operational lifespans. This underscores the need for careful material selection that prioritizes not only high sputtering resistance but also durability under RF sheath-enhanced sputtering environments.

Moreover, the study emphasizes the critical role of incorporating realistic RF sheath potentials and high-fidelity ion energy-angle distributions into erosion modeling. These factors are essential for accurately predicting material lifetimes and optimizing PFC designs. For fusion reactors with extensive RF heating systems, capturing the effects of RF sheath rectification with precision will be crucial for informed decisions on PFC design, proactive maintenance scheduling, and operational planning. By integrating these insights into future designs, fusion experiments can better mitigate material erosion challenges and improve overall reactor efficiency and reliability.

\section{Summary}
\label{sec:8}

This study highlights the critical influence of RF-induced sheaths on W erosion at ICRH antenna limiters in WEST tokamak, demonstrating that RF sheath potentials significantly enhance erosion rates compared to thermal sheath conditions. Quantitatively, integrated W erosion flux under RF sheath conditions during the ICRH phase was \(4.35 \times 10^{17}\) particles/s, over an order of magnitude higher than \(3.29 \times 10^{16}\) particles/s under thermal sheath conditions in the same phase, and more than 30 times greater than the \(1.39 \times 10^{16}\) particles/s observed under thermal sheath conditions during the ohmic phase. These results emphasize the dominant role of RF sheath potentials in driving material erosion.

High-charge-state oxygen ions (\(\rm O^{6^+}\), \(\rm O^{7^+}\), and \(\rm O^{8^+}\)) are identified as the primary contributors to W sputtering, with \(\rm O^{8^+}\) alone accounting for approximately 52\% of the total erosion flux under RF sheath conditions. In contrast, contributions from $\rm D^+$ ions (\(\rm D^+\)) are negligible across all scenarios, underscoring the importance of ion charge state and energy in determining sputtering efficiency. Spatially, erosion hot spots are localized to areas of peak RF sheath potentials, as identified through COMSOL simulations, with erosion concentrated on the upper regions of the limiters and reduced near the OMP where magnetic flux lines are tangential to the surface.

The STRIPE framework integrates advanced tools, including SolEdge3x, COMSOL, RustBCA and GITR, to model these complex PMI. Validation against spectroscopic measurements of neutral W emissions (W-I 400.9 nm) for WEST discharge \#57877 demonstrates STRIPE’s ability to capture overall trends in erosion fluxes. However, some discrepancies are observed between modeled and experimental data at specific spatial locations, particularly in regions where plasma conditions or localized sheath effects may not have been fully captured by the model. These differences highlight areas for further refinement in the representation of localized plasma dynamics and sheath potentials.

Despite these spatial disagreements, STRIPE provides critical insights into the mechanisms driving W erosion in RF environments. The framework’s capacity to model high-charge-state ion contributions, localized erosion hot spots, and RF sheath effects in 3D establishes it as a valuable tool for understanding PMI. These findings contribute to the broader goal of mitigating material erosion and impurity generation in fusion devices, providing key guidance for the design and operation of next-generation fusion reactors.

\section*{Acknowledgement}
 This material is based upon work supported by the U.S. Department of Energy, Office of Science, Office of Advanced Scientific Computing Research and Office of Fusion Energy Science, Scientific Discovery through Advanced Computing (SciDAC) program. This research used resources of the Fusion Energy Division, FFESD and the ORNL Research Cloud Infrastructure at the Oak Ridge National Laboratory, which is supported by the Office of Science of the U.S. Department of Energy under Contract No. DE-AC05-00OR22725.
\section*{References}
\bibliographystyle{ieeetr}  
\bibliography{west}
\end{document}